\documentclass[aps,twocolumn,showpacs]{revtex4}
\usepackage{amsmath}
\usepackage{epsfig}
\usepackage{booktabs}
\usepackage{multirow}
\usepackage{lipsum}
\usepackage{graphicx}

\begin{document}

\title{Strong decays of singly heavy baryons}

\author{Xiaoning Xie$^1$}\email[E-mail: ]{221002020@njnu.edu.cn}
\author{Juntao Tong$^1$}\email[E-mail: ]{191002007@njnu.edu.cn}
\author{Qi Huang$^1$}\email[E-mail: ]{06289@njnu.edu.cn}
\author{Hongxia Huang$^1$}\email[E-mail: ]{hxhuang@njnu.edu.cn (Corresponding author)}
\author{Jialun Ping$^1$}\email[E-mail: ]{jlping@njnu.edu.cn (Corresponding author)}
\affiliation{$^1$Department of Physics, Nanjing Normal University, Nanjing, Jiangsu 210097, China}

\begin{abstract}
More and more excited baryons have been reported experimentally, but many properties are still unclear. This work attempts to simultaneously study the masses and strong decay widths of some
singly heavy baryons, in order to provide possible quantum numbers for these states. The chiral quark model and the $^{3}P_{0}$ decay model are employed to calculate
the masses and decay widths of $\Lambda_{c(b)}$ and $\Sigma_{c(b)}$ baryons for all quantum numbers with $2S$, $1P$, and $2P$ waves. We considered not only two-body strong decays but also the influence of three-body decays. Our calculations show that: (i) For states with experimentally determined quantum numbers, such as $\Lambda_c(2595)$, $\Lambda_c(2625)$, $\Lambda_b(5912)$ and $\Lambda_b(5920)$, the results are consistent with experimental data and the conclusions of most theoretical studies. (ii) For states whose quantum numbers have not yet been fully determined experimentally, we provide possible interpretations. For example, our calculations tend to interpret $\Lambda_c(2910)$ is interpreted as a $J^P=\frac{3}{2}^-$ state with 1P-wave $\rho$-mode or a $J^P=\frac{1}{2}^-$ state with 2P wave $\lambda$-mode. $\Lambda_c(2940)$ can be interpreted as the $J^P=\frac{3}{2}^-$ state with 2P-wave $\lambda$-mode. For $\Lambda_c(2860)$, we offer a different interpretation, proposing that its mass and width closely match those of a 2P-wave $J^P=\frac{1}{2}^{-}$ state. It is hoped that our calculations can provide valuable information for the experimental and theoretical studies of heavy baryons.

%Since more and more heavy baryons have been discovered experimentally recently, the prediction of their quantum numbers has piqued our interest, and we try to study the experimentally discovered $\Lambda_{c(b)}$ and $\Sigma_{c(b)}$ baryons from both energy spectrum and decay perspectives. The energy spectra and decay widths of the $\Lambda_{c(b)}$ and $\Sigma_{c(b)}$ baryons of the 2S, 1P and 2P waves for the full quantum number are considered in our calculations. In calculating the energy spectrum, we apply the chiral quark model to obtain the energy spectrum of the excited state of the P-wave by means of experimentally determined baryon fitting parameters for the ground state. And in the study of decays, we introduce two correction factors in the transition operator to reduce the mass translation, we consider the effects of both two-body strong decays and three-body decays. Finally, we compare the results with the states that have been found experimentally to give our predictions for the quantum number.

%In our calculations it is argued that the recent discovery of $\Lambda_c(2910)$ by the Belle Collaboration can be interpreted as a $J^P=\frac{3}{2}^{-}$ state of the 1P-wave $\rho$ mode or a $J^P=\frac{1}{2}^{-}$ state of the 2P-wave and $\Lambda_c(2940)$ can be interpreted as the $J^P=\frac{3}{2}^{-}$ state of a 2P wave.
\end{abstract}

\maketitle

\setcounter{totalnumber}{5}

\section{\label{sec:introduction}Introduction}

%Recently, Belle Collaboration reported a new excited $\Lambda_c$ state, namely $\Lambda_c(2910)$ via investigating the $\bar{B}^{0}\rightarrow \Sigma _{c}(2455)^{0}\pi ^{+} \bar{p} $ decay process~\cite{Belle:2022hnm}. The mass and width of $\Lambda_c(2910)$ are measured to be:
%\begin{eqnarray}
%m(\Lambda _{c}(2910)^{+})&=&2913.8\pm5.6\pm3.8 ~\mathrm{MeV} \\
%\Gamma(\Lambda _{c}(2910)^{+})&=&51.8\pm20.0\pm18.8 ~\mathrm{MeV}
%\end{eqnarray}
%Belle Collaboration thinks that this state is possibly a good candidate for $\Lambda_c(\frac{1}{2},2P)$.

In recent years, with the ground state heavy baryons well established, the experimental focus has been on discovering more and more excited heavy baryons. For $\Lambda_c$,
at present, seven excited states of $\Lambda_c$ have been found experimentally: $\Lambda_c(2595)^+$, $\Lambda_c(2625)^+$, $\Lambda_c(2765)^+$, $\Lambda_c(2860)^+$, $\Lambda_c(2880)^+$, $\Lambda_c(2910)$, $\Lambda_c(2940)^+$.
As early as 1995, the CLEO Collaboration discovered two low excited states: $\Lambda_c(2595)$ and $\Lambda_c(2625)$~\cite{CLEO:1994oxm}. Similarly,
in 2000, the CLEO Collaboration found $\Lambda_c(2765)^+$ in the final state of $\Lambda_c^+\pi^+\pi^-$~\cite{CLEO:2000mbh}, but it is not clear whether it is $\Lambda_c$ or $\Sigma_c$.
In 2017, LHCb Collaboration reported a new resonance $\Lambda_c(2860)^+$ with a quantum number of $\frac{3}{2}^{+}$ in the $D^0p$ channel and the masses, widths and quantum numbers of $\Lambda_c(2880)^+$ and $\Lambda_c(2940)^+$ resonances were measured~\cite{LHCb:2017jym}.
In 2023, Belle Collaboration reported a new excited $\Lambda_c$ state, namely $\Lambda_c(2910)$ with a mass of $2913.8\pm5.6\pm3.8$MeV and a decay width of $51.8\pm20.0\pm18.8$MeV
via investigating the $\bar{B}^{0}\rightarrow \Sigma _{c}(2455)^{0}\pi ^{+} \bar{p} $ decay process~\cite{Belle:2022hnm}.
They identified that this state is possibly a good candidate for $\Lambda_c(\frac{1}{2},2P)$. The experimental observation process of the excited state of $\Lambda_b$ is similar
to that of $\Lambda_c$. First, two narrow $\Lambda_b$ states, $\Lambda_b(5912)$ and $\Lambda_b(5920)$ were found in the $\Lambda_b^0\pi^+\pi^-$ invariant mass spectrum
in the LHCb Collaboration~\cite{LHCb:2012kxf}. Subsequently, two highly excited states $\Lambda_b(6146)$ and $\Lambda_b(6152)$ were likewise discovered by LHCb Collaboration
in 2019~\cite{LHCb:2019soc}.
In 2020, the CMS Collaboration confirmed the existence of these two highly excited states and at the same time they discovered a new $\Lambda_b$ excited state $\Lambda_b(6072)$.
However, compared with $\Lambda_c$ and $\Lambda_b$, experiments have not found much for the excited states of $\Sigma_c$ and $\Sigma_b$,
with only $\Sigma_c(2800)$ and $\Sigma_b(6097)$ reported so far ~\cite{Belle:2004zjl, LHCb:2018haf}.
The mass, width and decay modes of all $\Lambda_c$, $\Lambda_b$, $\Sigma_c$, $\Sigma_b$ found experimentally are listed in Table~\ref{1-1}.
As new heavy baryons are discovered, many theoretical models and methods are used to study them,
including lattice QCD~\cite{Padmanath:2013bla, Bahtiyar:2015sga, Perez-Rubio:2015zqb, Bahtiyar:2016dom},
the heavy hadron chiral perturbation theory~\cite{Cheng:2006dk, Jiang:2015xqa, Cheng:2015naa, Kawakami:2019hpp},
QCD sum rules~\cite{Chen:2016phw, Wang:2017vtv, Aliev:2018lcs, Cui:2019dzj, Azizi:2020tgh, Xin:2023usd,Agaev:2017ywp, Chen:2017sci, Chen:2017gnu, Aliev:2018vye, Yang:2023fsc, Tan:2023opd},
relativistic flux tube model~\cite{Chen:2014nyo, Jakhad:2023mni},
effective Lagrangian approach~\cite{Huang:2018bed},
quark model~\cite{Karliner:2015ema, Thakkar:2016dna, Nagahiro:2016nsx, Chen:2018vuc, Yao:2018jmc, Yang:2017qan, Yang:2019lsg, Faustov:2020gun, Ortiz-Pacheco:2020hmj, Kakadiya:2022zvy, Kumakawa:2021ujz, Yu:2022ymb, Li:2022xtj, Pan:2023hwt,Gandhi:2019xfw, Gandhi:2019bju, Kakadiya:2021jtv},
quark-diquark approach~\cite{Torcato:2023ijg}, chiral effective theory~\cite{Kim:2020imk}, $^{3}P_{0}$ model~\cite{Ye:2017dra, Ye:2017yvl, Gandhi:2018lez, Guo:2019ytq, Liang:2019aag, Liang:2020kvn, Gong:2021jkb, Yu:2023bxn, Garcia-Tecocoatzi:2022zrf, Luo:2023sra}, and so on.

\begin{table*}[hbt]
\caption{The experimental information of the excited $\Lambda_c$, $\Lambda_b$, $\Sigma_c$, $\Sigma_b$ baryons, from the PDG ~\cite{ParticleDataGroup:2024cfk}.\label{1-1}}
\setlength{\tabcolsep}{14pt}
\resizebox{\textwidth}{!}{
\begin{tabular}{ccccccccc}
\hline \hline
&\multicolumn{5}{c}{Experimental information}~ \\
&States~ &$J^P$~ &Mass(Mev)~ &Width(Mev)~ &Decay channels~ \\ \hline
&$\Lambda _{c}(2595)^+$~ &$\frac{1}{2}^-$~ &$2592.25\pm0.28$~ &$2.59\pm0.30\pm0.47$~ &$\Lambda _{c}^{+}\pi^{+}\pi^{-},\Sigma _{c}(2455)^{++}\pi^{-},\Sigma _{c}(2455)^{0}\pi^{+}$~ \\
&$\Lambda _{c}(2625)^+$~ &$\frac{3}{2}^-$~ &$2628.00\pm0.15$~ &$<0.52$~ &$\Lambda _{c}^{+}\pi^{+}\pi^{-},\Sigma _{c}(2455)^{++}\pi^{-},\Sigma _{c}(2455)^{0}\pi^{+}$~ \\
&$\Lambda _{c}(2765)^+$~ &$?^?$~ &$2766.6\pm2.4$~ &50~ &$\Lambda _{c}^{+}\pi^{+}\pi^{-}$~ \\
&$\Lambda _{c}(2860)^+$~ &$\frac{3}{2}^+$~ &$2856.1_{-1.7}^{+2.0}\pm 0.5 _{-5.6}^{+1.1} $~ &$67.6_{-8.1}^{+10.1}\pm 1.4 _{-20.0}^{+5.9}$~ &$D^0p$~ \\
&$\Lambda _{c}(2880)^+$~ &$\frac{5}{2}^+$~ &$2881.63\pm0.24$~ &$5.6_{-0.6}^{+0.8}$~ &$\Lambda _{c}^{+}\pi^{+}\pi^{-},\Sigma _{c}(2455)^{0,++}\pi^{\pm},\Sigma _{c}(2520)^{0,++}\pi^{\pm},D^0p$~ \\
&$\Lambda _{c}(2910)^+$~ &$?^?$~ &$2913.8\pm3.8$~ &$51.8\pm20.0\pm18.8$~ &$\Sigma _{c}(2455)^{++}\pi^{-},\Sigma _{c}(2455)^{0}\pi^{+}$~\\
&$\Lambda _{c}(2940)^+$~ &$\frac{3}{2}^-$~ &$2939.6_{-1.5}^{+1.3}$~ &$20_{-5}^{+6}$~ &$\Sigma _{c}(2455)^{0,++}\pi^{\pm},D^0p$~ \\
&$\Sigma _{c}(2800)^{++} $~ &$?^?$~ &$2801_{-6}^{+4}$~ &$75_{-13}^{+18}$~ &$\Lambda _{c}^{+}\pi$~\\
&$\Sigma _{c}(2800)^{+} $~ &$?^?$~ &$2792_{-5}^{+14}$~ &$62_{-23}^{+37}$~ &$\Lambda _{c}^{+}\pi$~\\
&$\Sigma _{c}(2800)^{0} $~ &$?^?$~ &$2806_{-7}^{+5}$~ &$72_{-15}^{+22}$~ &$\Lambda _{c}^{+}\pi$~\\
&$\Lambda _{b}(5912)^0$~ &$\frac{1}{2}^-$~ &$5912.19\pm0.17$~ &$<0.25$~ &$\Lambda _{b}^{0}\pi^{+}\pi^{-}$~ \\
&$\Lambda _{b}(5920)^0$~ &$\frac{3}{2}^-$~ &$5920.09\pm0.17$~ &$<0.19$~ &$\Lambda _{b}^{0}\pi^{+}\pi^{-}$~ \\
&$\Lambda _{b}(6070)^0$~ &$\frac{1}{2}^+$~ &$6072.3\pm2.9\pm0.2$~ &$72\pm11\pm2$~ &$\Lambda _{b}^{0}\pi^{+}\pi^{-}$~ \\
&$\Lambda _{b}(6146)^0$~ &$\frac{3}{2}^+$~ &$6146.2\pm 0.4$~ &$2.9\pm 1.3\pm0.3$~  &$\Lambda _{b}^{0}\pi^{+}\pi^{-}$~ \\
&$\Lambda _{b}(6152)^0$~ &$\frac{5}{2}^+$~ &$6152.5\pm0.4$~ &$2.1\pm 0.8\pm0.3$~ &$\Lambda _{b}^{0}\pi^{+}\pi^{-}$~ \\
&$\Sigma _{b}(6097)^{+} $~ &$?^?$~ &$6095.8\pm1.7\pm0.4$~ &$31.0\pm5.5\pm0.7$~ &$\Lambda _{b}\pi^{+}$~\\
&$\Sigma _{b}(6097)^{+} $~ &$?^?$~ &$6098.0\pm1.7\pm0.5$~ &$28.9\pm4.2\pm0.9$~ &$\Lambda _{b}\pi^{-}$~\\
\hline \hline
\end{tabular}}
\end{table*}

At present, most of the theoretical work on heavy baryons focuses on energy spectrum and decay widths, which are generally discussed separately.
There has been a lot of work studying singly heavy baryons from the perspective of the energy spectrum. In Ref.~\cite{Yu:2022ymb},
the excited state energy spectra of $\Lambda_Q$, $\Sigma_Q$ and $\Omega_Q$ ($Q=c,d$) are calculated by ISG method in the framework of the relativistic quark model.
In their calculations, $\Lambda_c(2765)$ and $\Lambda_b(6070)$ are interpreted as 2S states of $\Lambda_c$ and $\Lambda_b$, respectively,
and $\Lambda_c(2940)$ is interpreted as 2P($\frac{1}{2}^-$) states of $\Lambda_c$.
Besides, this work points out that both $\Sigma_c(2800)$ and $\Sigma_b(6097)$ can be considered $P$-wave excited states,
but their quantum numbers cannot be confirmed based on calculations of the energy spectrum alone. The work of Ref.~\cite{Gandhi:2019xfw} also calculated the masses of excited states of nonstrange singly charmed baryons in hypercentral constituent quark model. Their conclusion for $\Lambda_c$ is the same as Ref.~\cite{Yu:2022ymb}, but for the quantum number of $\Sigma_c(2800)$,
they believe that it can be interpreted as a $J^P=\frac{1}{2}^-$ state of the $P$-wave.
A relativistic flux tube model also can be used to calculate the energy spectrum of singly charmed baryons~\cite{Jakhad:2023mni}. In their calculations,
the $\Lambda_c(2940)$ can be assigned to one of the 2P states with $J^P=\frac{1}{2}^-$ or $J^P=\frac{3}{2}^-$ and $\Sigma_c(2800)$ is a good candidate for $1P$ state with $J^P=\frac{3}{2}^-$.
%The author of Ref.~\cite{Wang:2023wii} used the QCD sum rule to calculate the D-wave energy spectrum of charmed baryons. As there are few highly excited states found in experiments, they mainly compare the results with the newly discovered $\Omega_c(3327)$ and assigned the $\Omega_c(3327)$ to be the D-wave state with the $J^P=\frac{3}{2}^+$.

In addition, there is a lot of work that has studied the heavy baryon from a decay width perspective with the $^{3}P_{0}$ model.
In Ref.~\cite{Guo:2019ytq}, the author calculated the $P$-wave, $D$-wave decay width of $\Lambda_{c}$. $\Lambda_{c}(2595)$ and $\Lambda_{c}(2625)$ are identified with
$1P$-wave $J^{P}=\frac{1}{2}^{-}$ and $J^{P}=\frac{3}{2}^{-}$, respectively. $\Lambda_{c}(2860)$ and $\Lambda_{c}(2880)$ may be $1D$-wave charmed baryons.
However, the author of Ref.~\cite{Gong:2021jkb} held a different theoretical interpretation of the $\Lambda_{c}(2880)$, which has also been interpreted
in terms of highly excited state decay as a $1F$ wave of $J^{P}=\frac{5}{2}^{-}$. There are also theorists who have plotted the variation of the singly baryon decay widths
with mass.~\cite{Wang:2022dmw}, and according to the results of the plot, the $\Lambda_c(2910)$ is close to the $\frac{5}{2}^-$ state of the $1P$ wave $\rho$-mode.
Moreover they found that the strong decay width is sensitive to changes in mass.
Light-cone QCD sum rules are also one of the commonly used methods for solving strong decay problems.
The work of Ref.~\cite{Yang:2021lce} considered two-body as well as three-body strong decays, interpreting $\Sigma_c(2800)$ as a mixed state of $P$-waves.

Much of the above work have studied heavy baryon excited states separately from the point of view of energy spectrum and decay width.
In our group's previous work we have studied baryons, tetraquark states, pentaquark states using the chiral quark model~\cite{Tan:2019knr,Yang:2019dxd,Hu:2021nvs,Qin:2020gxr},
and the strong decay of mesons using the $^{3}P_{0}$ model~\cite{Wang:2014voa,Chen:2014ztr,Chen:2017sjs}.
So it is interesting to calculate the energy spectrum and the decay width at the same time in the framework of the same model. The consistent calculation can provide a more reliable
explanation the excited states found in experiments and provide information on excited heavy baryons states for experimental searching.
The paper is organized as follows. In Section II, we give a brief review of ChQM, GEM and $^{3}P_{0}$ decay model.
In Section III , we study the strong decay behaviors of the S-wave, 2S-wave, P-wave and 2P-wave $\Lambda_c$, $\Lambda_b$, $\Sigma_c$, $\Sigma_b$ baryons.
Finally, we give a brief summary of this work in the last section.

\section{THEORETICAL FRAMEWORK}
In our work, we apply the Gaussian expansion method in the framework of the chiral quark model to compute the singly heavy baryon energy spectrum and bring the computed masses and wave functions
into the $^{3}P_{0}$ model to compute the strong decay widths of the states. In addition to considering the two-body decays, three-body decay is also taken into account.

\subsection{Chiral quark model(ChQM)}
Chiral quark model is one of the most common approaches to describe hadron spectra, hadron-hadron interactions and multiquark states. The quark interaction potential in the model is composed of central force, tensor force and spin-orbit coupling force. In this model, in addition to one-gluon exchange (OGE), the massive constituent quarks also interact with each other through Goldstone boson exchange. Besides, the color confinement and the scalar $\sigma$ meson (chiral partner, acting on $u$ and $d$ quark only) exchange are also introduced.
The Hamiltonian of ChQM is given as follows:
\begin{eqnarray}
H  &=  &\sum_{i=1}^{n}\left( m_i+\frac{p^2_i}{2m_i}\right)-T_{CM}
      + \sum_{j>i=1}^{n}[ V_{CON}({{\bf r}_{ij}})\nonumber\\
    &&+V_{OGE}({{\bf r}_{ij}})+V_{\chi}({{\bf r}_{ij}})+V_s({{\bf r}_{ij}})],
\end{eqnarray}
where $m_i$ is the constituent mass of quark(antiquark), $p_i$ is momentum of quark and $T_{CM}$ is the kinetic energy of the center-of mass motion.

There are no colored particles in nature, and there are no free quarks, they are always imprisoned inside the hadron, which is the characteristic of color confinement, so the model phenomenally introduces the confinement potential:
\begin{eqnarray}
V_{CON}^C(\mathbf r_{ij})&=& [-a_c(1-e^{-\mu_cr}+ \Delta)] \boldsymbol{\lambda^c_i} \cdot \boldsymbol{\lambda^c_j}\\
V_{CON}^{SO}(\mathbf r_{ij})&=&-\boldsymbol{\lambda^c_i} \cdot \boldsymbol{\lambda^c_j}\frac{a_c\mu_ce^{-\mu cr_{ij}}}{4m_i^2m_j^2r_{ij}}[((m_i^2+m_j^2)(1-2a_s)\nonumber\\
&&+4m_im_j(1-a_s))
 \times(\mathbf S_+\cdot\mathbf L)\nonumber\\
&&+(m_j^2-m_i^2)(1-2a_s)(\mathbf S_+\cdot\mathbf L)]
\end{eqnarray}
where $a_c$, $\mu_c$, $\Delta$ are model parameters, $\boldsymbol{\lambda^c_i}$ represent the SU(3) color Gell-Mann matrices, $\boldsymbol S\cdot \boldsymbol L$ is spin-orbit coupling operator

In addition to the spontaneous breaking of chiral symmetry, the model also introduces the single gluon exchange potential (OGE), in which the central force, tensor force and
spin-orbit coupling force form as follows:
\begin{eqnarray}
     V_{oge}^C(\mathbf r_{ij})&=&\frac{1}{4}\alpha_{s_{q_{i}q_{j}}} \boldsymbol{\lambda^c_i} \cdot \boldsymbol{\lambda^c_j}[ \frac{1}{ r_{ij}}- \frac{1}{6m_im_j}\nonumber\\
     &&(\boldsymbol{\sigma}_i \cdot\boldsymbol{\sigma}_j)\frac{e^{-\frac{r_{ij}}{r_0{\mu}}}}{r_{ij}r_0(\mu)^2} ]  \\
     V_{oge}^T(\mathbf r_{ij})&=&-\frac{1}{16} \frac{\alpha_{s_{q_{i}q_{j}}}}{m_im_j} \boldsymbol{\lambda^c_i} \cdot \boldsymbol{\lambda^c_j} [ \frac{1}{r_{ij}^3}- \frac{e^{-r_{ij}/r_g(\mu)}}{r_{ij}}\nonumber\\
&&(\frac{1}{r_{ij}^2} +\frac{1}{3r_{g}^2(\mu)}+\frac{1}{r_{ij}r_g(\mu)}) ]S_{ij}\\
     V_{oge}^{SO}(\mathbf r_{ij})&=&-\frac{1}{16} \frac{\alpha_{s_{q_{i}q_{j}}}}{m_i^2m_j^2} \boldsymbol{\lambda^c_i} \cdot \boldsymbol{\lambda^c_j} [ \frac{1}{r_{ij}^3}- \frac{e^{-r_{ij}/r_g(\mu)}}{r_{ij}^3}\nonumber\\
     &&(1+\frac{r_{ij}}{r_g(\mu)} )]\times [((m_i+m_j)^2 +2m_im_j)\nonumber\\
     &&(\mathbf{S}_+ \cdot \mathbf{L} )+(m_j^2-m_i^2)(\mathbf{S}_- \cdot \mathbf{L} ) ]
\end{eqnarray}
where $S_{ij}$ is the quark tensor operator, $S_{ij}=3(\mathbf{\sigma_i} \cdot  \vec{r}_{ij})(\mathbf{\sigma_j} \cdot \vec{r}_{ij})-\mathbf{\sigma_i} \cdot \mathbf{\sigma_j}$.

Due to chiral symmetry spontaneous breaking, Goldstone boson exchange potentials appear between light quarks:
\begin{eqnarray}
V_{\pi}(\mathbf{r}_{ij})&=&\frac{g_{ch}^2}{4\pi}\frac{m_{\pi}^2}{12m_im_j}\frac{\Lambda_{\pi}^2}{\Lambda_{\pi}^2-m_{\pi}^2}m_\pi\sum_{a=1}^3 \lambda_i^a \lambda_j^a \nonumber\\
&&\left \{  (\boldsymbol{\sigma}_i \cdot\boldsymbol{\sigma}_j)
\left[ Y(m_\pi (\mathbf{r}_{ij}))-\frac{\Lambda_{\pi}^3}{m_{\pi}^3}Y(\Lambda_{\pi} (\mathbf{r}_{ij})) \right] \right. \nonumber\\
&&\left.+\left[ H(m_\pi \mathbf r_{ij})-\frac{\Lambda_{\pi}^3}{m_{\pi}^3}H(\Lambda_{\pi} \mathbf r_{ij}) \right]S_{ij} \right \}
\end{eqnarray}
\begin{eqnarray}
V_{K}(\mathbf{r}_{ij})&=&\frac{g_{ch}^2}{4\pi}\frac{m_{K}^2}{12m_im_j}\frac{\Lambda_{K}^2}{\Lambda_{K}^2-m_{K}^2}m_K\sum_{a=4}^7 \lambda_i^a \lambda_j^a \nonumber\\
&&\left \{  (\boldsymbol{\sigma}_i \cdot\boldsymbol{\sigma}_j)
\left[ Y(m_K (\mathbf{r}_{ij}))-\frac{\Lambda_{K}^3}{m_{K}^3}Y(\Lambda_{K} (\mathbf{r}_{ij})) \right] \right. \nonumber\\
&&\left.+\left[ H(m_K \mathbf r_{ij})-\frac{\Lambda_{K}^3}{m_{K}^3}H(\Lambda_{K} \mathbf r_{ij}) \right]S_{ij} \right \}
\end{eqnarray}
\begin{eqnarray}
V_{\eta}(\mathbf{r}_{ij})&=&\frac{g_{ch}^2}{4\pi}\frac{m_{\eta}^2}{12m_im_j}\frac{\Lambda_{\eta}^2}{\Lambda_{\eta}^2-m_{\eta}^2}m_\eta\nonumber\\
&&\left[\lambda_i^8 \lambda_j^8 \cos\theta_P
 - \lambda_i^0 \lambda_j^0 \sin \theta_P \right]\nonumber\\
&&\left \{  (\boldsymbol{\sigma}_i \cdot\boldsymbol{\sigma}_j)
\left[ Y(m_\eta (\mathbf{r}_{ij}))-\frac{\Lambda_{\eta}^3}{m_{\eta}^3}Y(\Lambda_{\eta} (\mathbf{r}_{ij})) \right] \right. \nonumber\\
&&\left.+\left[ H(m_\eta \mathbf r_{ij})-\frac{\Lambda_{\eta}^3}{m_{\eta}^3}H(\Lambda_{\eta} \mathbf r_{ij}) \right]S_{ij} \right \}\\
H(x)&=&(1+\frac{3}{x}+\frac{3}{x^2})Y(x),~~~~ Y(x)=e^{-x}/x
\end{eqnarray}

Besides, the nonet (the extension of chiral partner $\sigma$ meson) exchange is also used in this work, which introduces other higher multi-pion terms that are simulated
through the full nonet of scalar mesons exchange between two constituent quarks~\cite{Garcilazo:2007ss}.
\begin{eqnarray}
V_{s}(\mathbf r_{ij}) &=& \upsilon_{a_0}(\mathbf r_{ij})\sum \limits _{a=1}^{3} \mathbf \lambda_i^a \cdot \mathbf\lambda_j^a + \upsilon_{\kappa}(\mathbf r_{ij})\sum\limits_{a=4}^{7} \mathbf\lambda_i^a \cdot \mathbf\lambda_j^a \\
&&+ \upsilon_{f0}(\mathbf r_{ij}) \mathbf\lambda_i^8 \cdot \mathbf\lambda_j^8 \nonumber\upsilon_{\sigma}(r_{ij}) \lambda_i^0 \cdot \lambda_j^0  \\
\upsilon_{s}(r_{ij})&=&-\frac{g_{ch}^2}{4\pi}\frac{\Lambda_s^2}{\Lambda_s^2-m_s^2}m_s\left[ Y(m_s r_{ij})-\frac{\Lambda_s}{m_s}Y(\Lambda_s r_{ij})\right]\nonumber\\
&&s=\sigma,a_0,\kappa,f_0
\end{eqnarray}

We fit the parameters by the baryon and meson masses of the ground state, which have been experimentally determined so far. Since it is difficult to fit both baryons and mesons
well with one set of parameters, different parameter values are given for the $\alpha_{s_{q_{i}q_{j}}}$ of mesons and baryons, with the previous one denoting baryons
and the later one denoting mesons. The model parameters are listed in Table ~\ref{2-1}, and the calculated baryon and meson masses are presented in the Table ~\ref{2-2}
with the experimental values.

\begin{table}[h]
\caption{Quark model parameters:$m_\pi=0.70$ fm$^{-1}$, $m_k=2.51$fm$^{-1}$, $m_\eta=2.77$fm$^{-1}$, $\Lambda_\pi=4.20$fm$^{-1}$, $\Lambda_k=\Lambda_\eta=5.20$fm$^{-1}$, $m_\sigma=3.42$fm$^{-1}$, $\Lambda_\sigma=4.20$fm$^{-1}$, $m_{a_0}=m_{f_0}=m_{\kappa}=4.97$fm$^{-1}$, $\Lambda_{a_0}=\Lambda_{f_0}=\Lambda_{\kappa}=5.20$fm$^{-1}$. $g^2_{ch}/(4\pi)=0.54$, $\theta_P=-15^\circ$}.
\label{2-1}
\begin{tabular}{c|cc cc} \hline \hline
                  &$m_u$=$m_d\;$(MeV)   &~~~~313\\
 Quark masses      &$m_s\;$(MeV)  &~~~~555\\
                    &$m_c\;$(MeV)  &~~~~1800\\
                     &$m_b\;$(MeV)  &~~~~5112\\  \hline
                   &$a_c$ (MeV)  &~~~~184.84\\
     Confinement       &$\mu_c$ (fm$^{-1})$  &~~~~0.683\\
                    &$\Delta$ (MeV)  &~~~~56.677\\
                    &$a_s$          &~~~~0.252\\   \hline
                 &$\hat{r}_0~$(MeV~fm)  &~~~~36.878\\
                 &$\hat{r}_g~$(MeV~fm)  &~~~~275.82\\
                    &$\alpha_{s_{uu}}$  &~~~~0.669/0.80\\
                    &$\alpha_{s_{us}}$  &~~~~0.7/0.724\\
                    &$\alpha_{s_{ss}}$  &~~~~0.707\\
    OGE    &$\alpha_{s_{uc}}$  &~~~~0.902/0.804\\
                   &$\alpha_{s_{sc}}$  &~~~~0.834\\
                   &$\alpha_{s_{ub}}$  &~~~~0.787\\
                   &$\alpha_{s_{sb}}$  &~~~~0.721\\\hline \hline

\end{tabular}
\end{table}

\begin{table}[h]
\caption{The masses of ground-state baryons and mesons(unit: MeV), Experimental information from PDG~\cite{ParticleDataGroup:2024cfk}}
\label{2-2}
\begin{tabular}{ccccccccc}
\hline \hline
      &$N$~  &$\Lambda$~ &$\Sigma$~ &$\Xi$~ &$\Delta$~  &$\Sigma^*$~ &$\Xi^*$~  &$\Omega$~ \\ \hline
ChQM~ &942~  &1154~      &1160~     &1343~  &1245~      &1374~       &1503~     &1653~  \\
Expt~ &939~  &1116~      &1193~     &1318~  &1232~      &1383~       &1533~     &1672~ \\ \hline

      &$\Lambda_c$~ &$\Sigma_c$~ &$\Xi_c$~ &$\Xi'_c$~ &$\Omega_c$~ &$\Sigma^*_c$~  &${\Xi'_c}^*$~ &$\Omega^*_c$  \\ \hline
ChQM~ &2282~        &2467~       &2472~    &2559~     &2694~       &2534~          &2640~         &2785~\\
Expt~ &2286~        &2455~       &2469~    &2579~     &2695~       &2520~          &2645~         &2770~ \\  \hline

      &$\Lambda_b$~ &$\Sigma_b$~ &$\Xi_b$~ &$\Xi'_b$~ &$\Omega_b$~ &$\Sigma^*_b$~  &${\Xi'_b}^*$~   \\ \hline
ChQM~ &5611~        &5826~       &5794~    &5920~     &6033~       &5848~          &5948~    \\
Expt~ &5620~        &5811~       &5797~    &5935~     &6045~       &5830~          &5950~     \\  \hline

      &$\pi$~  &$K$~      &$D$~  \\ \hline
ChQM~ &139~   &494~   &1869~  \\
Expt~ &140~   &494~   &1870~  \\
\hline \hline
\end{tabular}
\end{table}

\subsection{Gaussian expansion method(GEM)}
The Gaussian expansion method~\cite{Kamimura:1988zz} is to expand the orbital wave function of the system through a series of Gaussian wave functions with varying widths or centers,
then solve the Schr\"{o}dinger equation based on the variational principle, at last one arrives a generalized eigenproblem of matrix. After diagonalizing the matrix, one obtains
the eigenenergy and eigenfunction of the system.

We use a set of gaussians to expand the radial part of the orbital wave function which is shown below,
\begin{eqnarray}
\psi_{lm}(\mathbf{r})=\sum^{n_{max}}_{n=1}c_{nl}\phi^{G}_{nlm}(\mathbf{r})
\end{eqnarray}
\begin{eqnarray}
\phi^{G}_{nlm}(\mathbf{r})=\emph{N}_{nl}r^{l}e^{-\nu_{n}r^{2}}\emph{Y}_{lm}(\hat{\mathbf{r}})
\end{eqnarray}
where $N_{nl}$ is the normalization constant,
\begin{eqnarray}
\emph{N}_{nl}=\left(\frac{2^{l+2}(2\nu_{n})^{l+3/2}}{\sqrt\pi(2l+1)!!}\right)^{\frac{1}{2}},
\end{eqnarray}
and $c_{nl}$ is the variational parameter, which is determined by the dynamics of the system. The Gaussian size
parameters are chosen according to the following geometric progression:
\begin{eqnarray}
\nu_{n}=\frac{1}{r^{2}_{n}}, r_{n}=r_{min}a^{n-1}, a=\left(\frac{r_{max}}{r_{min}}\right)^{\frac{1}{n_{max}-1}},
\end{eqnarray}
where $n_{max}$ is the number of Gaussian functions, and $n_{max}$ is determined by the convergence of the results.In the present calculation, $n_{max}=8$.

\subsection{$^{3}P_{0}$ model}
$^{3}P_{0}$ model also known as quark pair generation model, was first proposed by Micu in 1969 ~\cite{Micu:1968mk} and further developed by Le Yaouanc, Ackleh, Roberts et al~\cite{LeYaouanc:1972vsx, LeYaouanc:1973ldf, LeYaouanc:1974cvx}. $^{3}P_{0}$ model is mainly used to study the two-body strong decay problem allowed by OZI(Okubo, Zweg and Iizuka) rule. This model can well describe the strong decay process of hadrons (baryons, mesons), so it is widely used in the study of the strong decay process of hadrons.

\begin{figure}[ht]
\begin{center}
\epsfxsize=3.0in \epsfbox{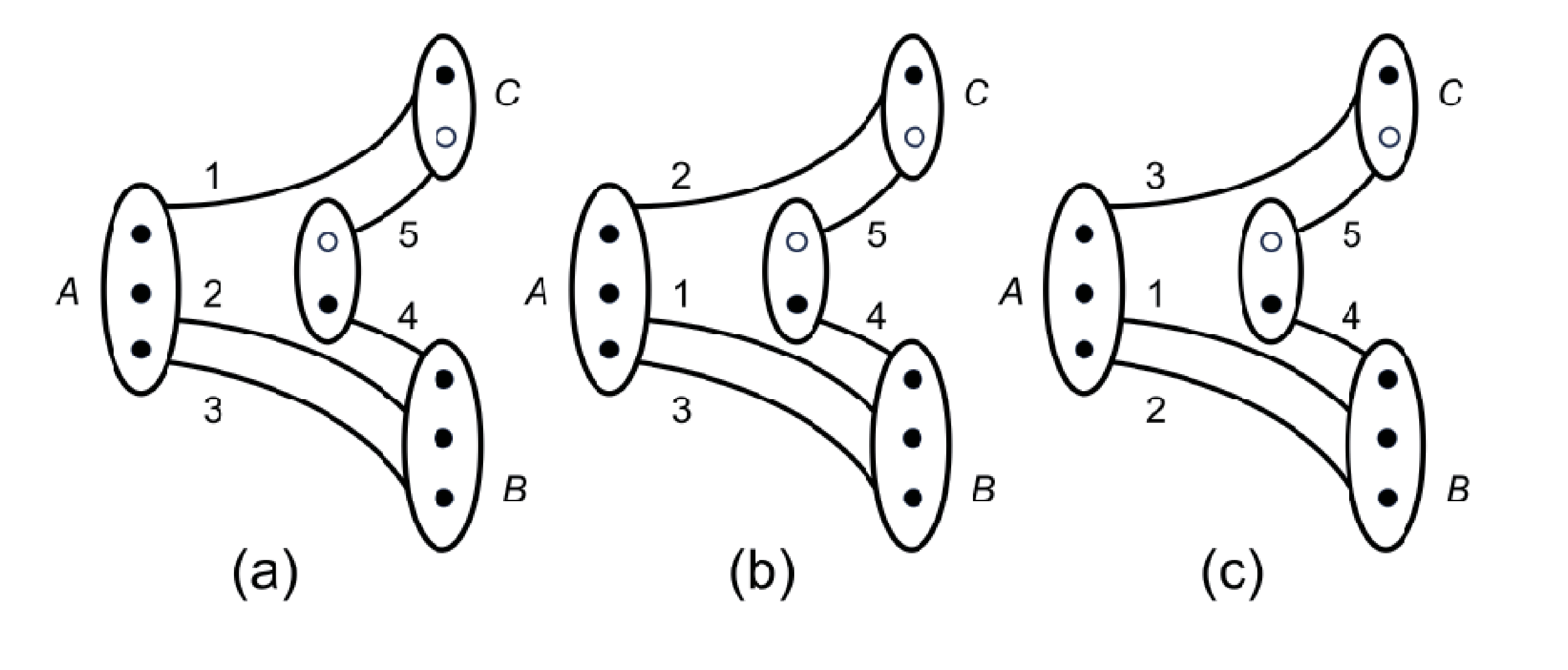} \vspace{-0.1in}
\caption{Baryon decay process of $A\rightarrow B+C$ in the $^{3}P_{0}$ model.\label{3p0}}
\end{center}
\end{figure}

According to this model, a pair of quarks with $J^{PC}=0^{++}$ is created from the vacuum when a hadron decays. During the quark rearrangement process, the new $q\bar{q}$ pair that forms from the vacuum combines with the $qqq$ (three quarks) within the initial baryon, resulting in the creation of the outgoing meson and baryon. When considering a baryon A, decaying to a baryon B and meson C, there are usually three types of recombination, as shown in FIG.\ref{3p0}.

In the $^{3}P_{0}$ model, the transition operator $T$ of the decay $A\rightarrow BC$ can be express as follows,
\begin{align}
T= & -3 \gamma \sum_m\langle 1 m 1-m \mid 00\rangle \int d^3 \boldsymbol{p}_4 d^3 \boldsymbol{p}_5 \delta^3\left(\boldsymbol{p}_4+\boldsymbol{p}_5\right) \notag\\
&\times \mathcal{Y}_1^m\left(\frac{\boldsymbol{p}_4-\boldsymbol{p}_5}{2}\right) \chi_{1-m}^{45} \phi_0^{45} \omega_0^{45} b_{4}^{\dagger}\left(\boldsymbol{p}_4\right) d_{4}^{\dagger}\left(\boldsymbol{p}_5\right),
\label{t1}
\end{align}
where $\gamma$ describes the probability for creating a quark-antiquark pair with momenta $\boldsymbol{p}_4$ and $\boldsymbol{p}_5$, respectively from the $0^{++}$ vacuum.
The solid harmonic polynomial $\mathcal{Y}_1^m(\boldsymbol{p})\equiv \left|p\right| Y_{1}^{m}(\theta{p},\phi{p})$ reflects the $P-$wave distribution of
the $q_{4}\bar{q}_{5}$ in the momentum space. $\phi_{0}^{45}=(u\bar{u}+d\bar{d}+s\bar{s})/ \sqrt{3}$, $\omega_{0}^{45}=(r\bar{r}+g\bar{g}+b\bar{b})/ \sqrt{3}$, and $\chi_{1-m}^{45}$ are the flavor , color, and spin wave functions of the $q_{4}\bar{q}_{5}$, respectively.

In our group's previous work~\cite{Chen:2017mug}, when we used the original transition operator in the momentum space Eq.(\ref{t1}) to calculate the mass shift problem of light mesons,
we found that these light mesons had relatively large mass shifts, which was very unreasonable. To solve the problem, the convergence factors $e^{-\frac{r^{2}}{4f^{2}}}$ and
the damping factor $e^{-\frac{R_{AV}^{2} }{R_{0}^{2}}}$ are introduced to modify the transition operator in the $^{3}P_{0}$ model~\cite{Chen:2017mug,Huang:2023jec}.

\begin{align}
T= & -3i \gamma \sum_m\langle 1 m 1-m \mid 00\rangle \int d^3 \boldsymbol{r}_4 d^3 \boldsymbol{r}_5\boldsymbol{r}(\frac{1}{2\pi })^{\frac{3}{2}}2^{-\frac{5}{2}}f^{-5}\notag \\ &Y_1^m(r) e^{-\frac{r^{2}}{4f^{2}}}e^{-\frac{R_{AV}^{2} }{R_{0}^{2}}} \chi_{1-m}^{45} \phi_0^{45} \omega_0^{45} b_{4}^{\dagger}\left(\boldsymbol{r}_4\right) d_{4}^{\dagger}\left(\boldsymbol{r}_5\right),
\label{t2}
\end{align}

The matrix element for the transition $A\rightarrow B+C$  can then be written:
\begin{align}
\langle B C|T| A\rangle=\delta^3\left(\boldsymbol{P}_A-\boldsymbol{P}_B-\boldsymbol{P}_C\right) \mathcal{M}^{M_{J_A} M_{J_B} M_{J_C}}.
\end{align}
where $\boldsymbol{P}_B$, $\boldsymbol{P}_C$ are the momenta of the $B$ and $C$ hadrons that appear in the final state, with $\boldsymbol{P}_A = \boldsymbol{P}_B + \boldsymbol{P}_C = 0$ in the center-of-mass frame of baryon $A$.
The helicity amplitude of the process  $A\rightarrow B+C$  in the center of mass frame of $A$ is
\begin{align}
 &\mathcal{M}^{M_{J_A} M_{J_B} M_{J_C}}(A \rightarrow B C)\notag\\
= &\sqrt{8 E_A E_B E_C} \gamma \sum_{\substack{M_{L_A}, M_{S_A},\\M_{L_B}, M_{S_B},\\M_{L_C}, M_{S_C},m}}\left\langle L_A M_{L_A} S_A M_{S_A} \mid J_A M_{J_A}\right\rangle\notag \\
&\times\left\langle L_B M_{L_B} S_B M_{S_B} \mid J_B M_{J_B}\right\rangle\left\langle L_C M_{L_C} S_C M_{S_C} \mid J_C M_{J_C}\right\rangle\notag\\
&\times\langle 1 m 1-m \mid 00\rangle\left\langle\chi_{S_C M_{S_C}}^{235} \chi_{S_B M_{S_B}}^{14} \mid \chi_{S_A M_{S_A}}^{123} \chi_{1-m}^{45}\right\rangle \notag\\
& \times\left\langle\varphi_C^{235} \varphi_B^{14} \mid \varphi_A^{123} \varphi_0^{45}\right\rangle I_{M_{L_B}, M_{L_C}}^{M_{L_A}, m}(\mathbf{p})
\end{align}
where the spatial integral $I_{M_{L_B}, M_{L_C}}^{M_{L_A}, m}(\mathbf{p})$ is defined as
\begin{align}
& I_{M_{L_B}, M_{L_C}}^{M_{L_A}, m}(\mathbf{p}) \notag\\
= & \int \mathrm{d}^3 \mathbf{k}_1 \mathrm{~d}^3 \mathbf{k}_2 \mathrm{~d}^3 \mathbf{k}_3 \mathrm{~d}^3 \mathbf{k}_4 \mathrm{~d}^3 \mathbf{k}_5 \delta^3\left(\mathbf{k}_{\mathbf{4}}+\mathbf{k}_{\mathbf{5}}\right) \notag\\
& \times \delta^3\left(\mathbf{k}_{\mathbf{1}}+\mathbf{k}_{\mathbf{2}}+\mathbf{k}_{\mathbf{3}}-\mathbf{P}_{\mathbf{A}}\right) \delta^3\left(\mathbf{k}_{\mathbf{1}}+\mathbf{k}_{\mathbf{4}}-\mathbf{P}_{\mathrm{B}}\right) \notag\\
& \times \delta^3\left(\mathbf{k}_{\mathbf{2}}+\mathbf{k}_{\mathbf{3}}+\mathbf{k}_{\mathbf{5}}-\mathbf{P}_{\mathbf{C}}\right) \notag\\
& \times \psi_{n_B L_B M_{L_B}}^*\left(\mathbf{k}_1, \mathbf{k}_4\right) \psi_{n_C L_C M_{L_C}}^*\left(\mathbf{k}_2, \mathbf{k}_3, \mathbf{k}_5\right) \notag\\
& \times \psi_{n_A L_A M_{L_A}}\left(\mathbf{k}_1, \mathbf{k}_2, \mathbf{k}_3\right) \mathcal{Y}_1^m\left(\frac{\mathbf{k}_{\mathbf{4}}-\mathbf{k}_{\mathbf{5}}}{2}\right) .
\end{align}

Here, $\psi_{nLM_{L}}$ is the (Fourier transform) of the wave function in Eq.(2), which is obtained via the self-consistent solution of the Hamiltonian problem in Eq.(1). $\left\langle\chi_{S_C M_{S_C}}^{235} \chi_{S_B M_{S_B}}^{14} \mid \chi_{S_A M_{S_A}}^{123} \chi_{1-m}^{45}\right\rangle$ and $\left\langle\varphi_C^{235} \varphi_B^{14} \mid \varphi_A^{123} \varphi_0^{45}\right\rangle$ denote the spin and flavor matrix element respectively.

In the $^{3}P_{0}$ model, the hadronic decay width $\Gamma$ of a process $A \rightarrow B + C$ is as follows,
\begin{align}
\Gamma =\pi^{2}\frac{\left|\mathbf{p}\right|}{M_{A}^{2}}\sum_{JL}\left|\mathcal{M}^{JL}\right|^{2},
\end{align}
where nonrelativistic phase-space is assumed,
\begin{align}
\left|\mathbf{p}\right|=\frac{\sqrt{[M_{A}^{2}-(M_{B}+M_{C})^{2}][M_{A}^{2}-(M_{B}+M_{C})^{2}]}}{2M_{A}},
\end{align}
with $M_{A}$, $M_{B}$, $M_{C}$ being the masses of the hadrons A, B, C; the partial wave amplitude $\mathcal{M}_{JL}(A\rightarrow BC)$ is related to the helicity amplitude via:
\begin{align}
& \mathcal{M}^{J L}(A \rightarrow B C)=\frac{\sqrt{2 L+1}}{2 J_A+1} \sum_{M_{J_B}, M_{J_C}}\left\langle L 0 J M_{J_A} \mid J_A M_{J_A}\right\rangle \notag\\
& \quad \times\left\langle J_B M_{J_B} J_C M_{J_C} \mid J M_{J_A}\right\rangle \mathcal{M}^{M_{J_A} M_{J_B} M_{J_C}}(\mathbf{P}).
\end{align}
where $J=J_{B}-J_{C}$, $J_{A}=J_{B}+J_{C}+L$, $M_{J_{A}}=M_{J_{B}}+M_{J_{C}}$.

In calculating the decay width with the $^{3}P_{0}$ model, there are a total of three parameters: $R_0$, $f$ and $\gamma$, parameters $R_0$ and $f$ are taken from Ref.~\cite{Chen:2017mug}, when $R_0=1.0fm$, $f=0.3fm$, it is possible to solve the problem of excessively large mass translations. For the $\gamma$ parameter we fitted the decay width of the ground state baryons $\Sigma_c$, $\Sigma_c^*$, $\Xi_c^*$, $\Sigma_b$, $\Sigma_b^*$, $\Xi_b^*$ and found that the obtained decay width is more reasonable when $\gamma=30.0$. The theoretical widths of the ground state baryons are shown in Table~\ref{2-3}. Using the above fitted parameters we will calculate the decay width of the heavy baryon excited states.

\begin{table}[h]
\caption{Decay widths(Mev)of S wave baryons}
\label{2-3}
\begin{tabular}{ccccccccc}
\hline \hline
& ~  &Channels~  &$\Gamma_{th}$~ &$\Gamma_{exp}$~\\
\hline
&$\Sigma_c(2455)$~ &$\Lambda_c \pi$~  &3.54~ &1.83-2.3~\\
&$\Sigma_c^*(2520)$~ &$\Lambda_c \pi$~  &22.15~ &14.78-17.2~\\
&$\Xi_c^*(2645)$~ &$\Xi_c \pi$~ &0.92~ &2.14-2.35~ \\
&$\Sigma_b(5810)$~ &$\Lambda_b \pi$~  &8.78~  &5.0-5.3~\\
&$\Sigma_b^*(5830)$~ &$\Lambda_b \pi$~  &14.90~  &9.4-10.4~\\
&$\Xi_b^*(5955)$~ &$\Xi_b \pi$~ &0.24~ &0.9-1.65~ \\
\hline \hline
\end{tabular}
\end{table}

\subsection{Three-body decay}

From PDG, we can see that the decay modes of particles observed experimentally include not only two-body strong decay, but also three-body strong decay, radiative decay and weak decay. In this paper, we not only consider two-body decay, but also consider the process of three-body decay process. The three-body decay is shown in FIG.\ref{3b1}:
\begin{figure}[ht]
\begin{center}
\includegraphics[height=4cm,width=8cm]{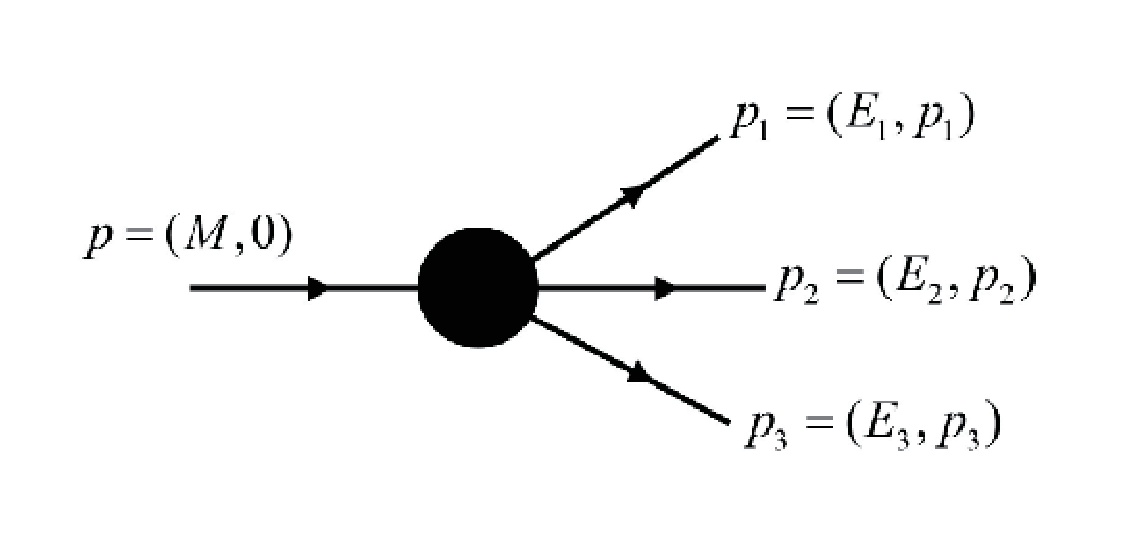} \vspace{-0.8cm}
\caption{The three-body decay process.\label{3b1}}
\end{center}
\end{figure}

In the rest system of the initial particle, the momentum $\boldsymbol{p}_i$ of the three final particles must be in the same plane: $\boldsymbol{p}_1+\boldsymbol{p}_2+\boldsymbol{p}_3=0$. The momentum space direction of the parent particle can be determined by three Euler angles$(\alpha ,\beta ,\gamma)$:
\begin{eqnarray}
d\Gamma=\frac{1}{(2\pi )^{5}}\frac{1}{16M}|\mathcal{M}|^2dE_1dE_2d\alpha d(\cos\beta)d\gamma
\end{eqnarray}
If the rest system of 1, 2 particles is used, then:
\begin{eqnarray}
d\Gamma=\frac{1}{(2\pi)^{5}}\frac{1}{16M^2}|\mathcal{M}|^2dm_{12}|\boldsymbol{p}_{1}^{*}||\boldsymbol{p}_3|d\Omega _{1}^{*}d\Omega_3
\end{eqnarray}
Where $\Omega_3$ is the momentum and solid Angle of a particle in the 1, 2 rest system, $|\boldsymbol{p}_{1}^{*}|$ and $|\boldsymbol{p}_3|$ are expressed as follows:
\begin{align}
|\boldsymbol{p}_{1}^{*}|&=\frac{1}{2m_{12}}[(m_{12}^2-(m_{12}+m_3)^2)(m_{12}^2-(m_{12}-m_3)^2)]^{1/2}\nonumber\\
|\boldsymbol{p}_3|&=\frac{1}{M}[(M^2-(m_{12}+m_3)^2)(M^2-(m_{12}-m_3)^2)]^{1/2}
\end{align}

When the initial particle is a scalar particle, integrating the Angle or averaging the spin yields the differential of the decay width:
\begin{align}
d\Gamma &=\frac{1}{(2\pi )^{3}}\frac{1}{8M}|\mathcal{M}|^2dE_1dE_2\\ \nonumber
&=\frac{1}{(2\pi )^{3}}\frac{1}{8M}|\mathcal{M}|^2\frac{dm_{12}^{2}dm_{23}^{2}}{4M^2}
\end{align}

\begin{figure}[ht]
\begin{center}
\includegraphics[height=4cm,width=6cm]{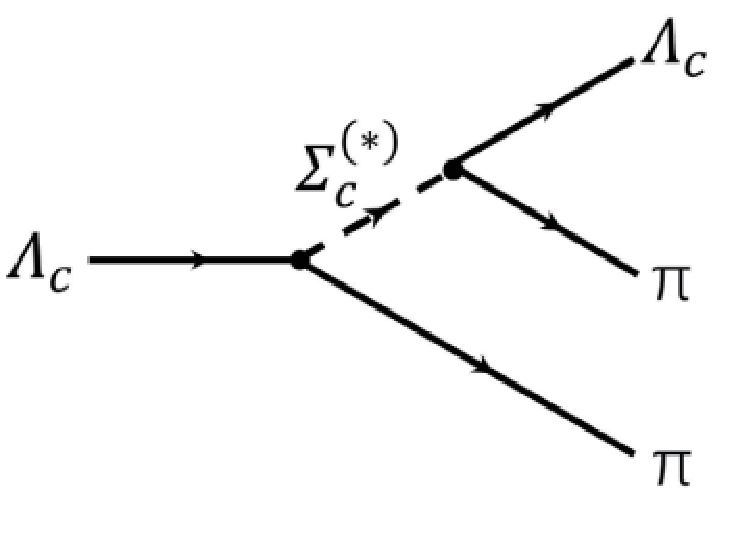} \vspace{-0.8cm}
\caption{Three body continuous decay process.\label{3b2}}
\end{center}
\end{figure}

In theory, it is difficult to directly solve the three-body strong decay, so we consider transforming the three-body process into two two-body decay processes. Taking $\Lambda_c(2765)$ as an example, firstly, we consider that $\Lambda_c(2765)$ decays to $\Sigma_c \pi$ and $\Sigma_c^* \pi$. The ground state $\Sigma_c$ and $\Sigma_c^*$ can continue to decay, and the final state of decay is $\Lambda_c \pi$, as shown in Fig\ref{3b2}. According to the above two processes, we can get the three-body decay of $\Lambda_c(2765) \to \Lambda_c \pi \pi$.

This decay mode is called successive decay ($A \to B+C,B\to D+F$), and its decay width is:

\begin{align}
&d\Gamma=\frac{M_{DF}^{2}}{M_{A}^{2}}\frac{|p|}{4\pi^2}\frac{\tilde{\Gamma}_{\small{B\to D+F }}\mathcal{|M_{A\to{B+C}}}|^2 }{2M_{B}^{2}[(M_{DF}-M_{B})^2+\tilde{\Gamma}_{\small{B\to D+F }}^2/4]}dM_{DF}\\
&\left|\mathbf{p}\right|=\frac{\sqrt{[M_{A}^{2}-(M_C+M_{DF})^2][M_{A}^{2}-(M_C-M_{DF})^2]}}{2M_A}
\end{align}
The upper and lower limits of the integral are $M_D+M_F$ to $M_A-M_C$, so we only need to get the decay width of the $B\to D+F$ process and the partial wave amplitude of the $A \to B+C$ process to get the decay width of the successive decay, which is approximately compared with the experiment as the result of three-body decay.

\section{Results and discussions}
In this work, we use the chiral quark model to calculate the masses of baryon excited states and apply the modified $^{3}P_{0}$ model to calculate the $2S$, $1P$ and $2P$-wave
singly heavy baryon decay width.

For the excited state of $P$ wave, there are two excitation modes: $\lambda$ mode and $\rho$ mode, we use the j-j coupling method, which has the relation that $\vec{L_{\lambda}}+\vec{L_{\rho}}=\vec{L}$, $\vec{L}+\vec{S_{\rho}}=\vec{J_{l}}$, $\vec{J_{l}}+\vec{S_Q}=\vec{J}$, $n_{\rho}$, $L_{\rho}$, $S_{\rho}$ denote the radial quantum number,
the orbital angular momentum, and the spin of the two light, while $n_{\lambda}$, $L_{\lambda}$ denote the nodal quantum number and the orbital angular momentum between the heavy quark
and the two-light-quark system. $L$ is the total orbital angular momentum of the baryon and $J$ denotes the total angular momentum.

\subsection{$\Lambda_c$}

For $\Lambda_c$ systems, the quantum numbers of 1P-wave and 2S-wave are shown in Table ~\ref{3-1}, where $\Lambda_{c1}(\frac{1}{2}^{-})$ denotes $j_l=1$, $J^P=\frac{1}{2}^{-}$ and $\tilde{\Lambda}_{c0}(\frac{1}{2}^-)$ denotes $\rho$-mode excitation, $j_l=0$, $J^P=\frac{1}{2}^{-}$. The theoretical mass and width of $\Lambda_c$ are shown in Table~\ref{3-2}. Considering that the mass will affect the results of the decay to some extent, we correct the theoretical mass to the experimental mass and to compare the values of the decay widths in Table ~\ref{3-3}, \ref{3-4}, \ref{3-5}.

$\mathit{\Lambda_c(2595)}$ and $\mathit{\Lambda_c(2625)}$: $\Lambda_c(2595)$ and $\Lambda_c(2625)$ were observed in $\Lambda _{c}^{+}\pi^{+}\pi^{-} $ channel by the CLEO Collaboration ~\cite{CLEO:1994oxm}. $\Lambda_c(2595)$ and $\Lambda_c(2625)$ had been experimentally observed to decay to $\Lambda _{c}^{+}\pi^{+}\pi^{-} $, $\Sigma_{c}(2455)^{++}\pi^{-}$ and $\Sigma_{c}(2455)^{0}\pi^{+}$. The branching fractions $\Gamma (\Lambda _{c}(2595)^{+}\to \Sigma_{c}(2455)^{++}\pi^{-})/\Gamma _{total}=24\pm 7\%$ , $\Gamma (\Lambda _{c}(2595)^{+}\to \Sigma_{c}(2455)^{0}\pi^{+})/\Gamma _{total}=24\pm 7\%$ and $\Lambda _{c}^{+}\pi^{+}\pi^{-} $ three-body decay with a fraction of about $18\pm 10\%$. The experimentally measured decay width of $\Lambda_c(2625)^+$ is less than 0.52 MeV~\cite{CLEO:1994oxm}, in which three-body decay still accounts for $67\%$, so the proportion of two-body decay $\Lambda_{c}(2625)^{+}\to \Sigma_{c}(2455)\pi$ is very small: $\Gamma(\Lambda_{c}(2625)^{+}\to \Sigma_{c}(2455)^{++}\pi^{-})/\Gamma _{total}<5\% $, $\Gamma(\Lambda_{c}(2625)^{+}\to \Sigma_{c}(2455)^{0}\pi^{+})/\Gamma _{total}<5\% $.

\begin{table}[h]
\caption{The quantum number of 1S 2S and 1P-wave $\Lambda_c$ baryons}
\label{3-1}
\begin{tabular}{cccccccccc}
\hline \hline
&Assignments~ &$J$~ &$J_l$~ &$n_\lambda$~ &$L_\rho$~ &$L_\lambda$~ &$L$~ &$S_\rho$~ \\ \hline
&$\Lambda_c(\frac{1}{2}^+,1S)$~ &$\frac{1}{2}$~ &0~ &0~ &0~ &0~ &0~ &0~ \\
&$\Lambda_c(\frac{1}{2}^+,2S)$~ &$\frac{1}{2}$~ &0~ &1~ &0~ &0~ &0~ &0~ \\ \hline
&$\Lambda_{c1}(\frac{1}{2}^-,1P)$~ &$\frac{1}{2}$~ &1~ &0~ &0~ &1~ &1~ &0~ \\
&$\Lambda_{c1}(\frac{3}{2}^-,1P)$~ &$\frac{3}{2}$~ &1~ &0~ &0~ &1~ &1~ &0~ \\
&$\tilde{\Lambda}_{c0}(\frac{1}{2}^-,1P)$~ &$\frac{1}{2}$~ &0~ &0~ &1~ &0~ &1~ &1~ \\
&$\tilde{\Lambda}_{c1}(\frac{1}{2}^-,1P)$~ &$\frac{1}{2}$~ &1~ &0~ &1~ &0~ &1~ &1~ \\
&$\tilde{\Lambda}_{c1}(\frac{3}{2}^-,1P)$~ &$\frac{3}{2}$~ &1~ &0~ &1~ &0~ &1~ &1~ \\
&$\tilde{\Lambda}_{c2}(\frac{3}{2}^-,1P)$~ &$\frac{3}{2}$~ &2~ &0~ &1~ &0~ &1~ &1~ \\
&$\tilde{\Lambda}_{c2}(\frac{5}{2}^-,1P)$~ &$\frac{5}{2}$~ &2~ &0~ &1~ &0~ &1~ &1~ \\
\hline \hline
\end{tabular}
\end{table}

\begin{table}[h]
\caption{Masses and decay widths of 2S and 1P-wave $\Lambda_c$ states. unit: MeV}
\label{3-2}
\begin{tabular}{ccccccccc}
\hline \hline
&$\Lambda_{c}$~ &Mass~ &$\Sigma_c\pi$~ &$\Sigma_c^*\pi$~ &ND~ &$\Lambda _{c}\pi\pi$~ &$\Gamma_{total}$~\\ \hline
&$\Lambda_c(\frac{1}{2}^+,2S)$~ &2709~ &9.22~ &3.90~ &$\setminus$~ &0.05~ &13.17~\\ \hline
&$\Lambda_{c1}(\frac{1}{2}^-,1P)$~ &2592~  &$\setminus$~ &$\setminus$~ &$\setminus$~ &$\setminus$~ &$\setminus$~\\
&$\tilde{\Lambda}_{c0}(\frac{1}{2}^-,1P)$~ &2796~ &$\setminus$~ &$\setminus$~ &$\setminus$~ &$\approx 0$~ &$\approx 0$~\\
&$\tilde{\Lambda}_{c1}(\frac{1}{2}^-,1P)$~ &2772~ &553.13~ &2.09~ &$\setminus$~ &2.19~ &557.41~\\
&$\Lambda_{c1}(\frac{3}{2}^-,1P)$~ &2619~ &0.003~ &$\setminus$~ &$\setminus$~ &$\approx 0$~ &0.003~\\
&$\tilde{\Lambda}_{c1}(\frac{3}{2}^-,1P)$~ &2807~ &10.13~ &509.93~ &$\setminus$~ &2.32~ &522.38~\\
&$\tilde{\Lambda}_{c2}(\frac{3}{2}^-,1P)$~ &2878~ &46.94~ &20.50~ &$\approx 0$~ &0.26~ &67.70~\\
&$\tilde{\Lambda}_{c2}(\frac{5}{2}^-,1P)$~ &2918~ &29.98~ &55.10~ &$\approx 0$~ &0.33~ &85.41~\\
\hline \hline
\end{tabular}
\end{table}

%$\Lambda_c(2595)$,$\Lambda_c(2625)$
In most theoretical work~\cite{Garcia-Tecocoatzi:2022zrf, Yu:2023bxn,Guo:2019ytq}, $\Lambda_c(2595)$ and $\Lambda_c(2625)$ are identified as 1P-wave excited states and form a doublet $\Lambda_{c1}(\frac{1}{2}^{-},\frac{3}{2}^{-})$.
From Table\ref{3-2} it can be seen that the theoretical masses of $\Lambda_{c1}(\frac{1}{2}^-)$ is very close to the experimental values of $\Lambda_c(2595)$ and the decay channels observed experimentally are $\Lambda_c\pi\pi$ and $\Sigma_c\pi$. Due to the theoretical mass of $\Lambda_c$ is lower than the sum of the theoretical masses of $\Sigma_c$ and $\pi$ in our calculation, the decay width of $\Lambda_c(2595)$ cannot be obtained by using theoretical mass, so in Table\ref{3-3} we use the experimental mass 2592 MeV to calculate the decay width . The total decay width obtained is as follows: the partial width of $\Sigma_c\pi$ is 1.39 MeV, which is well reproduced the partial width of $\Sigma_c\pi$ measured in experiment. Therefore, $\Lambda_c(2595)$ could be identified as a 1P-wave state with $J^P=\frac{1}{2}^-$.

\begin{table}[h]
\centering
\caption{Masses and decay widths of $\Lambda_c(2595)$ as 2S and 1P-wave states. unit: MeV}
\label{3-3}
\begin{tabular}{ccccccccc}
\hline \hline
&$\Lambda_{c}$~ &$Mass(Mev)$~ &$\Sigma_c\pi$~ &$\Sigma_c^*\pi$~ &ND~ &$\Lambda _{c}\pi\pi$~ &$\Gamma_{total}$~\\ \hline
&$\Lambda_c(\frac{1}{2}^+,2S)$~ &2592~ &$\approx$ 0.0~ &$\setminus$~ &$\setminus$~ &$\approx$ 0.0~ &$\approx$ 0.0~\\ \hline
&$\Lambda_{c1}(\frac{1}{2}^-,1P)$~ &2592~ &1.39~  &$\setminus$~  &$\setminus$~ &0.02~ &1.41~\\
&$\tilde{\Lambda}_{c0}(\frac{1}{2}^-,1P)$~ &2592~ &$\setminus$~ &$\setminus$~ &$\setminus$~ &$\setminus$~ &$\setminus$~\\
&$\tilde{\Lambda}_{c1}(\frac{1}{2}^-,1P)$~ &2592~ &6.14~ &$\setminus$~ &$\setminus$~ &0.08~ &6.22~\\
&$\Lambda_{c1}(\frac{3}{2}^-,1P)$~ &2592~ &$\approx$ 0.0~ &$\setminus$~ &$\setminus$~ &$\approx$ 0.0~ &$\approx$ 0.0~\\
&$\tilde{\Lambda}_{c1}(\frac{3}{2}^-,1P)$~ &2592~ &$\approx$ 0.0~ &$\setminus$~ &$\setminus$~ &$\approx$ 0.0~ &$\approx$ 0.0~\\
&$\tilde{\Lambda}_{c2}(\frac{3}{2}^-,1P)$~ &2592~ &$\approx$ 0.0~ &$\setminus$~ &$\setminus$~ &$\approx$ 0.0~ &$\approx$ 0.0~\\
&$\tilde{\Lambda}_{c2}(\frac{5}{2}^-,1P)$~ &2592~ &$\approx$ 0.0~ &$\setminus$~ &$\setminus$~ &$\approx$ 0.0~ &$\approx$ 0.0~\\
\hline \hline
\end{tabular}
\end{table}

From Table\ref{3-2} it can be seen that the theoretical masses of $\Lambda_{c1}(\frac{3}{2}^-)$ and is very close to the experimental values of $\Lambda_c(2625)$ and the decay width of $\Lambda_{c1}(\frac{3}{2}^-)$ is 0.003 MeV, approaching 0 MeV. When the experimental mass 2628 MeV is used to calculate the decay width of $\Lambda_c(2625)^+$, the result of $\Lambda_{c1}(\frac{3}{2}^-)$ obtained is 0.05 MeV in Table~\ref{3-4}, which is closer to the experimental width. Here, it is reasonable to treat $\Lambda_{c1}(\frac{3}{2}^{-})$ as $\Lambda_c(2625)$, which is consistent with the conclusions of most theoretical work.

\begin{table}[h]
\centering
\caption{Masses and decay widths of $\Lambda_c(2625)$ as 2S and 1P-wave states. unit: MeV}
\label{3-4}
\begin{tabular}{ccccccccc}
\hline \hline
&$\Lambda_{c}$~ &$Mass$~ &$\Sigma_c\pi$~ &$\Sigma_c^*\pi$~ &ND~ &$\Lambda _{c}\pi\pi$~ &$\Gamma_{total}$~\\ \hline
&$\Lambda_c(\frac{1}{2}^+,2S)$~ &2628~ &1.43~ &$\setminus$~ &$\setminus$~ &0.01~ &1.44~\\ \hline
&$\Lambda_{c1}(\frac{1}{2}^-,1P)$~ &2628~  &56.53~ &$\setminus$~ &$\setminus$~ &0.22~ &56.75~\\
&$\tilde{\Lambda}_{c0}(\frac{1}{2}^-,1P)$~ &2628~ &$\setminus$~ &$\setminus$~ &$\setminus$~ &$\setminus$~ &$\setminus$~\\
&$\tilde{\Lambda}_{c1}(\frac{1}{2}^-,1P)$~ &2628~ &247.90~ &$\setminus$~ &$\setminus$~ &0.98~ &248.88~\\
&$\Lambda_{c1}(\frac{3}{2}^-,1P)$~ &2628~ &0.05~ &$\setminus$~ &$\setminus$~ &$\approx$ 0.0~ &0.05~\\
&$\tilde{\Lambda}_{c1}(\frac{3}{2}^-,1P)$~ &2628~ &0.05~ &$\setminus$~ &$\setminus$~ &$\approx$ 0.0~ &0.05~\\
&$\tilde{\Lambda}_{c2}(\frac{3}{2}^-,1P)$~ &2628~ &0.09~ &$\setminus$~ &$\setminus$~ &$\approx$ 0.0~ &0.09~\\
&$\tilde{\Lambda}_{c2}(\frac{5}{2}^-,1P)$~ &2628~ &0.04~ &$\setminus$~ &$\setminus$~ &$\approx$ 0.0~ &0.04~\\
\hline \hline
\end{tabular}
\end{table}

%Lambda_c(2765)
$\mathit{\Lambda_c(2765)}$: In 2001, the CLEO Collaboration found $\Lambda_c(2765)$ in the final decay state of $\Lambda_c^+\pi^+\pi^-$ with a mass of $2766.6\pm2.4$ MeV and a decay width of about 50 MeV~\cite{CLEO:2000mbh}. The $I(J^P)$ quantum number of $\Lambda_c(2765)$ has not yet been experimentally determined, and it is still inconclusive whether $\Lambda_c(2765)$ is a $\Lambda_c$ or $\Sigma_c$ state or a overlapping states. In Ref.~\cite{Yu:2023bxn}, $^{3}P_{0}$ model is also used to calculate the decay width. They believe that $\Lambda_c(2765)$ can be interpreted as a $\Lambda_c(2S)$ state with a mass of 2764 MeV and a width of 25.34 MeV. However in Ref.~\cite{Garcia-Tecocoatzi:2022zrf}, their theoretical work did not find a suitable $\Lambda_c$ state to explain $\Lambda_c(2765)$. The author of Ref.~\cite{Guo:2019ytq} considers that $\Lambda_c(2765)$ cannot be considered as the P-wave state of $\Lambda_c$, but it may be 2S-wave or 1D-wave state. The result in Table ~\ref{3-2} show that the mass of the $\rho$-mode $\tilde{\Lambda}_{c0}(\frac{1}{2}^-)$, $\tilde{\Lambda}_{c1}(\frac{1}{2}^-)$, $\tilde{\Lambda}_{c1}(\frac{3}{2}^-)$ are close to that of $\Lambda_c(2765)$, but the decay width of $\tilde{\Lambda}_{c1}(\frac{1}{2}^-)$ and $\tilde{\Lambda}_{c1}(\frac{3}{2}^-)$ are too large by more than 500 MeV and must be excluded. $\tilde{\Lambda}_{c2}(\frac{3}{2}^-)$'s decay width 67.44 MeV is very close to the experimental width, but its mass of 2878 MeV exceeds the experimental width about 100 MeV. Thus it is unreasonable to describe $\Lambda_c(2765)$ as a P-wave $\Lambda_c$ state. However the theoretical mass 2709 MeV and width 13.12 MeV of the 2S-wave state are smaller than the experimental values. Considering that the mass will affect the results of the decay to some extent, we used the experimental masses to calculate the decay widths as in Table~\ref{3-5}. The decay width of $\Lambda_c$ 2S-wave states is 51.31 MeV, which is approach to the experimental width. $\Lambda_c(2765)$ can be interpreted as a $\Lambda_c$ 2S-wave state when we take into account the effect of mass on the decay widths.

\begin{table}[h]
\centering
\caption{Masses and decay widths of $\Lambda_c(2765)$ as 2S and 1P-wave states. unit: MeV}
\label{3-5}
\begin{tabular}{ccccccccc}
\hline \hline
&$\Lambda_{c}$~ &$Mass$~ &$\Sigma_c\pi$~ &$\Sigma_c^*\pi$~ &ND~ &$\Lambda _{c}\pi\pi$~ &$\Gamma_{total}$~\\ \hline
&$\Lambda_c(2S)$~ &2767~ &22.59~ &28.72~ &$\setminus$~ &0.20~ &51.51~\\ \hline
&$\Lambda_{c1}(\frac{1}{2}^-)$~ &2767~  &129.83~ &3.09~ &$\setminus$~ &0.53~ &133.45~\\
&$\tilde{\Lambda}_{c0}(\frac{1}{2}^-)$~ &2767~ &$\setminus$~ &$\setminus$~ &$\setminus$~ &$\setminus$~\\
&$\tilde{\Lambda}_{c1}(\frac{1}{2}^-)$~ &2767~ &556.19~ &2.86~ &$\setminus$~ &2.23~ &561.28~\\
&$\Lambda_{c1}(\frac{3}{2}^-)$~ &2767~ &6.85~ &103.90~ &$\setminus$~ &0.42~ &111.17~\\
&$\tilde{\Lambda}_{c1}(\frac{3}{2}^-)$~ &2767~ &6.43~ &469.72~ &$\setminus$~ &1.81~ &477.96~\\
&$\tilde{\Lambda}_{c2}(\frac{3}{2}^-)$~ &2767~ &11.67~ &2.74~ &$\setminus$~ &0.06~ &14.47~\\
&$\tilde{\Lambda}_{c2}(\frac{5}{2}^-)$~ &2767~ &5.09~ &4.29~ &$\setminus$~ &0.04~ &9.42~\\
\hline \hline
\end{tabular}
\end{table}

%Lambda_c(2910),Lambda_c(2940)
$\mathit{\Lambda_c(2910)}$ : In 2023, Belle Collaboration reported a new excited $\Lambda_c$ state, namely $\Lambda_c(2910)$ with a mass of $2913.8\pm5.6\pm3.8$ MeV and a decay width of $51.8\pm20.0\pm18.8$ MeV via investigating the $\bar{B}^{0}\rightarrow \Sigma _{c}(2455)^{0}\pi ^{+} \bar{p} $ decay process~\cite{Belle:2022hnm}. The experiment suggested that $\Lambda_c(2910)$ may be a 2P-wave state with $J^P=\frac{1}{2}^-$. The author of Ref. ~\cite{Yang:2023fsc} used the QCD sum rules to well interpret $\Lambda_c(2910)$ as a $J^P=\frac{1}{2}^-$ 2P-wave state. However, Ref.~\cite{Wang:2022dmw} suggests that $\Lambda_c(2910)$ may be a $\tilde{\Lambda}_{c2}(\frac{5}{2}^-)$ state of the $\rho$-mode of 1P-wave. From Table~\ref{3-2}, we can see that the mass and decay width of $\tilde{\Lambda}_{c2}(\frac{5}{2}^-)$ is 2918 MeV and 85.41 MeV, respectively, which is very close to experimental values. The case of 2P-wave is likewise under consideration in our work which shown in Table~\ref{3-6}. 2P-wave mass of the five $\rho$-modes are higher than 3000 MeV, which are unreasonable as $\Lambda_c(2910)$. The two $\lambda$-modes have masses of 2857 MeV and 2871 MeV and decay widths of 68.915 MeV and 46.502 MeV. So it's reasonable to interpret $\Lambda_c(2910)$ as $\Lambda_{c1}(\frac{1}{2}^-)$ 2P-wave state. Thus, we tend to interpret $\Lambda_c(2910)$ as a 1P-wave $\rho$-mode state with a quantum number of $J^P=\frac{5}{2}^-$, which is consistent with the results of Ref.~\cite{Wang:2022dmw}. Alternatively, it could also be described as a 2P-wave $\lambda$-mode $J^P=\frac{1}{2}^-$ excitation.

%It can be seen from the results in Table ~\ref{3-2}, the masses of $\tilde{\Lambda}_{c2}(\frac{3}{2}^-)$ and $\tilde{\Lambda}_{c2}(\frac{5}{2}^-)$ are 2878 MeV and 2918 MeV, respectively, and they decay to $\Sigma_c\pi$ and $\Sigma_c^*\pi$, with a total decay width of 67.704 MeV and 85.409 MeV. Our theoretical width of 1P-wave $\tilde{\Lambda}_{c2}(\frac{3}{2}^-)$ state is compatible with the experimental value.
 %The case of 2P-wave is likewise under consideration in our work which shown in Table~\ref{3-6}. 2P-wave mass of the five $\rho$-modes are higher than 3000 MeV, which are unreasonable as $\Lambda_c(2910)$. The two $\lambda$-modes have masses of 2857 MeV and 2871 MeV and decay widths of 68.915 MeV and 46.502 MeV. So it's reasonable to interpret $\Lambda_c(2910)$ as $\Lambda_{c1}(\frac{1}{2}^-)$ 2P-wave state.

$\mathit{\Lambda_c(2940)}$: In 2007, the Belle Collaboration discovered a new $\Lambda_c$ state $\Lambda_c(2940)$~\cite{Belle:2006xni}. The quantum number of $\Lambda_c(2940)$ is experimentally considered to be $\frac{3}{2}^{-}$, but the possibility of $\frac{1}{2}^-$ and $\frac{7}{2}^-$ cannot be ruled out. Similarly, many theorists have given their predictions for the various properties of $\Lambda_c(2940)$. In Ref.~\cite{Gong:2021jkb}, the mass spectra and decay widths of 1$D$, 2$P$ and 1$F$-wave states are calculated, and $\Lambda_c(2940)$ may be the excitation of 2P-wave $\frac{3}{2}^-$ state, $n_{\rho}=1$,$l_{\lambda}=1$. Ref.~\cite{Yu:2023bxn} suggests that $\Lambda_c(2940)$ is a 2P-wave $\frac{1}{2}^-$ state with a decay width of 12.40 MeV. There are also some work interpret it as the molecular state of $ND^*$~\cite{Yan:2023ttx}. Most theoretical work considered that $\Lambda_c(2940)$ is a 2P-wave state with ND as the main decay channel. Our work also considers the case of $\Lambda_c$ as a 2P-wave in Table~\ref{3-6}. In contrast, the mass 2871 MeV and decay width 46.50 MeV of $\Lambda_{c1}(\frac{3}{2}^-)$ are closer to the experimental values. Therefore, we tend to interpret $\Lambda_c(2940)$ as a $\frac{3}{2}^-$ state of a 2P-wave, which is consistent with the conclusions of Ref.~\cite{Yu:2023bxn}.

\begin{table}[h]
\centering
\caption{Masses and decay widths of 2P-wave $\Lambda_c$ states. unit: MeV}
\label{3-6}
\begin{tabular}{ccccccccc}
\hline \hline
&$\Lambda_{c}$~ &Mass~ &$\Sigma_c\pi$~ &$\Sigma_c^*\pi$~ &ND~ &$\Lambda _{c}\pi\pi$~ &$\Gamma_{total}$~\\ \hline
&$\Lambda_{c1}(\frac{1}{2}^-,2S)$~ &2857~  &11.43~ &0.12~ &57.32~ &0.05~ &68.92~\\
\hline
&$\tilde{\Lambda}_{c0}(\frac{1}{2}^-,1P)$~ &3058~ &$\approx 0$~ &$\approx 0$~ &$\approx 0$~ &$\approx 0$~ &$\approx 0$~\\
&$\tilde{\Lambda}_{c1}(\frac{1}{2}^-,1P)$~ &3042~ &92.36~ &0.01~ &$\approx 0$~ &0.37~ &92.74~\\
&$\Lambda_{c1}(\frac{3}{2}^-,1P)$~ &2871~ &0.120~ &5.89~ &40.47~ &0.02~ &46.50~\\
&$\tilde{\Lambda}_{c1}(\frac{3}{2}^-,1P)$~ &3064~ &4.09~ &91.45~ &$\approx 0$~ &0.37~ &95.91~\\
&$\tilde{\Lambda}_{c2}(\frac{3}{2}^-,1P)$~ &3112~ &26.14~ &9.71~ &$\approx 0$~ &0.14~ &35.99~\\
&$\tilde{\Lambda}_{c2}(\frac{5}{2}^-,1P)$~ &3140~ &20.40~ &36.33~ &$\approx 0$~ &0.22~ &56.95~\\
\hline \hline
\end{tabular}
\end{table}

%Lambda_c(2860),Lambda_c(2880)
$\mathit{\Lambda_c(2860)}$ and $\mathit{\Lambda_c(2880)}$: For $\Lambda_c(2860)^+$ and $\Lambda_c(2880)^+$, the LHCb Collaboration considers their quantum numbers to be $\frac{3}{2}^+$ and $\frac{5}{2}^+$, respectively. Many works interpret $\Lambda_c(2860)^+$ and $\Lambda_c(2880)^+$ as double states of the D-wave ~\cite{Faustov:2020gun, Yu:2022ymb, Garcia-Tecocoatzi:2022zrf} with quantum numbers $J^P=\frac{3}{2}^{+}$ and $J^P=\frac{5}{2}^{+}$. However, there are different interpretations for these two states, in Ref.~\cite{Chen:2017aqm}, The theoretical mass of the D-wave is very close to the experimental value, but the decay width of $\Lambda_c(2860)^+$ is much lower than the experimental value. the authors employ the $^{3}P_{0}$ model to calculate the decay width and prefer to interpret $\Lambda_c(2860)^+$ as the $J^P=\frac{3}{2}^{+}$ 1D-wave state and interpret $\Lambda_c(2880)^+$ as the $J^P=\frac{5}{2}^{-}$ 1F-wave state. In our work, the masses and decay widths of 1P-wave $\rho$-mode $\tilde{\Lambda}_{c2}(\frac{3}{2}^-)$ and $\tilde{\Lambda}_{c2}(\frac{5}{2}^-)$ are close to the experimental value of $\Lambda_c(2860)$, but the decay width of the ND channel is close to 0 MeV, which is not in accordance with the experimental value. From Table ~\ref{3-6}, the mass of $\Lambda_{c1}(\frac{1}{2}^-)$ 2P-wave state is 2857 MeV and the total width is 68.916 MeV, in which the main decay channel is ND, which is in perfect agreement with the experimental value, so we prefer to interpret $\Lambda_c(2860)$ as the 2P-wave state with quantum number $J^P=\frac{1}{2}^{-}$. For $\Lambda_c(2880)^+$, on the other hand, it has a very narrow decay width of 5.6 MeV, and we did not find any state that would explain it well in the calculations for 1P and 2P-wave , so it is difficult to explain $\Lambda_c(2880)^+$ well as a P-wave state in the present calculations, and $\Lambda_c(2880)^+$ may be a D-wave state, which will be explored in our subsequent work.

\subsection{$\Lambda_b$}

For $\Lambda_b$ systems, the quantum numbers of 1P-wave and 2S-wave are shown in Table ~\ref{3-1}, The theoretical mass and width of $\Lambda_b$ are shown in Table~\ref{3-2}.

\begin{table}[h]
\centering
\caption{The quantum number of 1S, 2S and 1P-wave $\Lambda_b$ baryons}
\label{3-7}
\begin{tabular}{cccccccccc}
\hline \hline
&Assignments~ &$J$~ &$J_l$~ &$n_\lambda$~ &$L_\rho$~ &$L_\lambda$~ &$L$~ &$S_\rho$~ \\ \hline
&$\Lambda_b(\frac{1}{2}^+,1S)$~ &$\frac{1}{2}$~ &0~ &0~ &0~ &0~ &0~ &0~ \\
&$\Lambda_b(\frac{1}{2}^+,2S)$~ &$\frac{1}{2}$~ &0~ &1~ &0~ &0~ &0~ &0~ \\ \hline
&$\Lambda_{b1}(\frac{1}{2}^+,1P)$~ &$\frac{1}{2}$~ &1~ &0~ &0~ &1~ &1~ &0~ \\
&$\Lambda_{b1}(\frac{3}{2}^+,1P)$~ &$\frac{3}{2}$~ &1~ &0~ &0~ &1~ &1~ &0~ \\
&$\tilde{\Lambda}_{b0}(\frac{1}{2}^-,1P)$~ &$\frac{1}{2}$~ &0~ &0~ &1~ &0~ &1~ &1~ \\
&$\tilde{\Lambda}_{b1}(\frac{1}{2}^-,1P)$~ &$\frac{1}{2}$~ &1~ &0~ &1~ &0~ &1~ &1~ \\
&$\tilde{\Lambda}_{b1}(\frac{3}{2}^-,1P)$~ &$\frac{3}{2}$~ &1~ &0~ &1~ &0~ &1~ &1~ \\
&$\tilde{\Lambda}_{c2}(\frac{3}{2}^-,1P)$~ &$\frac{3}{2}$~ &2~ &0~ &1~ &0~ &1~ &1~ \\
&$\tilde{\Lambda}_{b2}(\frac{5}{2}^-,1P)$~ &$\frac{5}{2}$~ &2~ &0~ &1~ &0~ &1~ &1~ \\
\hline \hline
\end{tabular}
\end{table}

$\mathit{\Lambda_b(5912)}$ and $\mathit{\Lambda_b(5920)}$: In 2012, the LHCb collaboration discovered two narrow states, $\Lambda_b(5912)$ and $\Lambda_b(5920)$, in the $\Lambda_b^0 \pi^+ \pi^-$ invariant mass spectrum~\cite{LHCb:2012kxf}, this was the first experimental report of excited states of the $\Lambda_b$. The decay widths of these two states are very narrow, both less than 1 MeV. In the majority of energy spectrum analyses concerning the $\Lambda_b(5912)$ and $\Lambda_b(5920)$, there is a consensus that these particles correspond to the $\frac{1}{2}^-$ and $\frac{3}{2}^-$ states of the 1P-wave $\lambda$-mode, respectively. From Table~\ref{3-8}, it can be seen that the masses of $\Lambda_{b1}(\frac{1}{2}^-)$ and $\Lambda_{b1}(\frac{3}{2}^-)$ states are 5905 MeV and 5915 MeV, respectively, which differ from the experimental values by no more than 10 MeV. The decay final states of both $\Lambda_b(5912)$ and $\Lambda_b(5920)$ are $\Lambda_b \pi \pi$. However, in our calculations for three-body decays, it is necessary to first consider $\Lambda_b$ decaying to $\Sigma_b \pi$, and then $\Sigma_b$ further decaying to $\Lambda_b \pi$. Since the threshold sum of $\Sigma_b \pi$ is higher than that of $\Lambda_b(5912)$ and $\Lambda_b(5920)$, our current calculations cannot obtain the three-body decay widths for $\Lambda_b(5912)$ and $\Lambda_b(5920)$. According to the analysis of energy spectrum, We prefer to interpret $\Lambda_b(5912)$ and $\Lambda_b(5920)$ as 1P-wave states with the quantum number $J^P=\frac{1}{2}^-$ and $J^P=\frac{3}{2}^-$, respectively.

\begin{table}[h]
\caption{Masses and decay widths of 2S and 1P-wave $\Lambda_c$ states. unit: MeV}
\label{3-8}
\begin{tabular}{ccccccccc}
\hline \hline
&$\Lambda_{b}$~ &Mass~ &$\Sigma_b\pi$~ &$\Sigma_b^*\pi$~ &$\Lambda _{b}\pi\pi$~  &$\Gamma_{total}$~\\ \hline
&$\Lambda_b(\frac{1}{2}^+,2S)$~ &6011~ &1.91~ &1.36~ &0.01~  &3.28~\\ \hline
&$\Lambda_{b1}(\frac{1}{2}^-,1P)$~ &5905~  &$\setminus$~ &$\setminus$~ &$\setminus$~  &$\setminus$~\\
&$\tilde{\Lambda}_{b0}(\frac{1}{2}^-,1P)$~ &6122~ &$\setminus$~  &$\setminus$~ &$\setminus$~ &$\setminus$~\\
&$\tilde{\Lambda}_{b1}(\frac{1}{2}^-,1P)$~ &6108~ &484.75~ &3.18~ &1.90~  &489.83~\\
&$\Lambda_{b1}(\frac{3}{2}^-,1P)$~ &5915~  &$\setminus$~ &$\setminus$~ &$\setminus$~  &$\setminus$~\\
&$\tilde{\Lambda}_{b1}(\frac{3}{2}^-,1P)$~ & 6128~ &4.27~ &476.40~ &1.84~  &482.51~\\
&$\tilde{\Lambda}_{b2}(\frac{3}{2}^-,1P)$~ &6210~ &29.01~ &21.73~ &0.20~  &50.94~\\
&$\tilde{\Lambda}_{b2}(\frac{5}{2}^-,1P)$~ &6235~ &17.16~ &46.87~ &0.25~  &64.28~\\
\hline \hline
\end{tabular}
\end{table}

$\mathit{\Lambda_b(6070)}$: The LHCb Collaboration observed a new $\Lambda_b$ state in the $\Lambda _{b}^{0}\pi^{+}\pi^{-}$ mass spectrum using a data sample of pp collisions in 2020~\cite{LHCb:2020lzx}, The mass and width of $\Lambda_b(6070)$ are measured to be:
\begin{eqnarray}
m(\Lambda _{b}(6070)^{+})&=&6072.3\pm2.9\pm0.6\pm0.2 ~\mathrm{MeV} \\
\Gamma(\Lambda _{c}(6070)^{+})&=&72\pm11\pm2 ~\mathrm{MeV}
\end{eqnarray}
The LHCb Collaboration considered that $\Lambda_b(6070)$ is consistent with the first radial excitation of the $\Lambda_b$ baryon: $\Lambda_b(2S)$. Some studies have examined the $\Lambda_b(6070)$ from the perspective of energy spectra and believe it can be interpreted as a 2S state~\cite{Yu:2022ymb,Wang:2020mxk}. Other works have investigated the $\Lambda_b(6070)$ from the decay perspective and found that the decay width of the 2S state is too small compared to the experimental value of $\Lambda_b(6070)$~\cite{Yu:2023bxn,Liang:2020kvn}.

%In Ref.~\cite{Yu:2022ymb}, the relativistic quark model is used to calculate the mass of the 2S-state of $\Lambda_b$ as 6041 MeV, and $\Lambda_b(6070)$ is approximated to be the 2S-wave $\Lambda_b$  state. In Ref.~\cite{Wang:2020mxk}, QCD sum rules are used to calculate the masses of $\Lambda_b$ 1S and 2S-wave states. They interpret $\Lambda_b(6070)$ as the 2S-wave state of the $\Lambda_b$. As for the calculation of the decay width of $\Lambda_b$, the total decay width of $\Lambda_b$ 2S state calculated in Ref~\cite{Yu:2023bxn} is 8.82 MeV, much lower than the experimental value of 72 MeV. In Ref~\cite{Liang:2020kvn}, the calculated decay width of 9.72 MeV of the 2S-wave state is also much lower than the experimental value, which is not very reasonable to explained $\Lambda_b(6070)$ as 2S-wave state. In their work, $\Lambda_b(6070)$ is also considered as a 1P-wave state, and their result show that $\Lambda_b(6070)$ is unlikely to be a pure 1P-wave state, but may be a mixed state.

\begin{table}[h]
\caption{Masses and decay widths of $\Lambda_b(6070)$ as 2S and 1P-wave states. unit: MeV}
\label{3-9}
\begin{tabular}{ccccccccc}
\hline \hline
&$\Lambda_{b}$~ &Mass~ &$\Sigma_b\pi$~ &$\Sigma_b^*\pi$~ &$\Lambda _{b}\pi\pi$~  &$\Gamma_{total}$~\\ \hline
&$\Lambda_b(\frac{1}{2}^+,2S)$~ &6072~ &11.70~ &18.89~ &0.12~  &30.71~\\ \hline
&$\Lambda_{b1}(\frac{1}{2}^-,1P)$~ &6072~  &110.21~ &2.48~ &0.44~  &113.13~\\
&$\tilde{\Lambda}_{b0}(\frac{1}{2}^-,1P)$~ &6072~ &$\setminus$~  &$\setminus$~ &$\setminus$~ &$\setminus$~\\
&$\tilde{\Lambda}_{b1}(\frac{1}{2}^-,1P)$~ &6072~ &456.73~ &1.84~ &1.81~  &460.38~\\
&$\Lambda_{b1}(\frac{3}{2}^-,1P)$~ &6072~  &$\setminus$~ &$\setminus$~ &0.39~  &0.39~\\
&$\tilde{\Lambda}_{b1}(\frac{3}{2}^-,1P)$~ &6072~ &1.66~ &413.34~ &1.62~  &416.62~\\
&$\tilde{\Lambda}_{b2}(\frac{3}{2}^-,1P)$~ &6072~ &3.06~ &1.73~ &0.02~  &4.81~\\
&$\tilde{\Lambda}_{b2}(\frac{5}{2}^-,1P)$~ &6072~ &1.35~ &2.69~ &0.02~  &4.06~\\
\hline \hline
\end{tabular}
\end{table}

Our theoretical masses and decay widths are listed in Table\ref{3-8}. For the 1P-wave state, although $\tilde{\Lambda}_{b1}(\frac{1}{2}^-)$ is the closest to the experimental mass, its decay width exceeds 400 MeV, which is obviously unreasonable. While $\tilde{\Lambda}_{b2}(\frac{3}{2}^-)$ and $\tilde{\Lambda}_{b2}(\frac{5}{2}^-)$ have a decay width of 50.94 MeV and 64.28 MeV, which are basically consistent with the experimental value, their mass is too high. Therefore, the result indicate that $\Lambda_b(6070)$ maybe not a P-wave state. For the 2S-wave state, our theoretical mass 6011 MeV and width 3.28 MeV are both lower than the experimental values. Considering that mass has a certain influence on the width, we recalculated the decay width of $\Lambda_b$ by using the experimental mass in Table~\ref{3-9}. Among them, the total decay width of 2S state is 30.71 MeV, although the decay width of the 2S state is still slightly lower than the experimental value, $\Lambda_b(6070)$ may be a 2S state from the perspective of combining the energy spectrum and decay.

 %which is still much smaller than the experimental value. Base on our theoretical prediction, it is unreasonable to interpret $\Lambda_b(6070)$ as 2S-wave state of $\Lambda_b$.

\subsection{$\Sigma_c$}

For the ground $\Sigma_c$ state, the masses, decay widths of $\frac{1}{2}^{+}$ and $\frac{3}{2}^{+}$ states and their quantum numbers have been experimentally determined. At present, only $\Sigma_c(2800)$ and $\Sigma_c(2765)$ excited states have been found in the experiment, among which $\Sigma_c(2765)$ cannot be determined whether it is a $\Lambda_c$ state or a $\Sigma_c$ state. For $\Sigma_{c}$, the quantum numbers and masses of the P-wave and 2S-wave states are listed in Table \ref{3-10}, The theoretical masses and decay widths are shown in Table \ref{3-11}.
\begin{table}[h]
\caption{The quantum number of 1S, 2S and 1P-wave $\Sigma_c$ baryons}
\label{3-10}
\begin{tabular}{cccccccccc}
\hline \hline
&Assignments~ &$J$~ &$J_l$~ &$n_\lambda$~ &$L_\rho$~ &$L_\lambda$~ &$L$~ &$S_\rho$~ ~\\ \hline
&$\Sigma_c(\frac{1}{2}^+,1S)$~ &$\frac{1}{2}$~ &0~ &0~ &0~ &0~ &0~ &0~ \\
&$\Sigma_c(\frac{3}{2}^+,1S)$~ &$\frac{3}{2}$~ &1~ &0~ &0~ &0~ &0~ &1~ \\
&$\Sigma_c(\frac{1}{2}^+,2S)$~ &$\frac{1}{2}$~ &0~ &1~ &0~ &0~ &0~ &0~ \\
&$\Sigma_c^*(\frac{3}{2}^+,2S)$~ &$\frac{3}{2}$~ &1~ &0~ &0~ &0~ &0~ &1~ \\ \hline
&$\Sigma_{c0}(\frac{1}{2}^-,1P)$~ &$\frac{1}{2}$~ &0~ &0~ &0~ &1~ &1~ &1~ \\
&$\Sigma_{c1}(\frac{1}{2}^-,1P)$~ &$\frac{1}{2}$~ &1~ &0~ &0~ &1~ &1~ &1~ \\
&$\Sigma_{c1}(\frac{3}{2}^-,1P)$~ &$\frac{3}{2}$~ &1~ &0~ &0~ &1~ &1~ &1~ \\
&$\Sigma_{c2}(\frac{3}{2}^-,1P)$~ &$\frac{3}{2}$~ &2~ &0~ &0~ &1~ &1~ &1~ \\
&$\Sigma_{c2}(\frac{5}{2}^-,1P)$~ &$\frac{5}{2}$~ &2~ &0~ &0~ &1~ &1~ &1~ \\
&$\tilde{\Sigma}_{c1}(\frac{1}{2}^-,1P)$~ &$\frac{1}{2}$~ &1~ &0~ &1~ &0~ &1~ &0~ \\
&$\tilde{\Sigma}_{c1}(\frac{3}{2}^-,1P)$~ &$\frac{3}{2}$~ &1~ &0~ &1~ &0~ &1~ &0~ \\
\hline \hline
\end{tabular}
\end{table}

\begin{table}[h]
\caption{Masses and decay widths of 2S and 1P-wave $\Sigma_c$ states. unit: MeV}
\label{3-11}
\begin{tabular}{ccccccccc}
\hline \hline
&$\Sigma_{cJ_l}(J^P)$~ &Mass~ &$\Lambda_c\pi$~ &$\Sigma_c\pi$~ &$\Sigma_c^*\pi$~ &$\Gamma_{total}$~ \\ \hline
&$\Sigma_{c}(\frac{1}{2}^+,2S)$~ &2845~ &0.98~ &29.71~ &15.67~ &46.36~ \\
&$\Sigma_{c}^*(\frac{3}{2}^+,2S)$~ &2878~ &3.17~ &1.67~ &19.37~ &24.23~ \\ \hline
&$\Sigma_{c0}(\frac{1}{2}^-,1P)$~ &2738~ &52.47~ &$\setminus$~ &$\setminus$~ &52.47~ \\
&$\Sigma_{c1}(\frac{1}{2}^-,1P)$~ &2746~ &$\setminus$~ &191.70~ &0.42~ &192.12~ \\
&$\Sigma_{c1}(\frac{3}{2}^-,1P)$~ &2767~ &$\approx$ 0.0~ &2.48~ &149.50~ &151.98~ \\
&$\Sigma_{c2}(\frac{3}{2}^-,1P)$~ &2781~ &32.48~ &5.92~ &1.34~ &39.74~ \\
&$\Sigma_{c2}(\frac{5}{2}^-,1P)$~ &2809~ &38.39~ &4.26~ &4.48~ &48.13~ \\
&$\tilde{\Sigma}_{c1}(\frac{1}{2}^-,1P)$~ &2839 &$\setminus$~ &162.29~ &15.28~ &177.49~ \\
&$\tilde{\Sigma}_{c1}(\frac{3}{2}^-,1P)$~ &2849~ &$\approx$ 0.0~ &24.80~ &181.09~ &205.89~ \\
\hline \hline
\end{tabular}
\end{table}

%Sigma_c(2800)
$\mathit{\Sigma_c(2800)}$: As early as 2005, $\Sigma_c(2800)$ was discovered in the invariant mass spectrum of $\Lambda_c \pi$ by Belle Collaboration~\cite{Belle:2004zjl}, and $\Sigma_c(2800)^{0}$ was observed in the Barbar Collaboration in 2008~\cite{BaBar:2008get}. Most theoretical results believe that $\Sigma_c(2800)$ is a 1P-wave $\lambda$-mode excited state. Similarly some theoretical work suggests $\Sigma_c(2800)$ maybe the mixing state of $\frac{3}{2}^{-}$ and $\frac{1}{2}^{-}$ or a pure $\frac{5}{2}^{-}$ state of the $\lambda$-mode 1P-wave. For example, in Ref.~\cite{Garcia-Tecocoatzi:2022zrf}, the energy spectrum and decay widths are calculated by using the constituent quark model and the $^3P_0$ model, and $\Sigma_c(2800)$ is considered to be $\frac{1}{2}^{-}$ of the 1P-wave state. Ref.~\cite{Yu:2023bxn} considers the mixing of the same quantum numbers and considers $\Sigma_c(2800)$ possibly to be a mixing state of $\left | \frac{3}{2}^{-},j=1\right \rangle $ and $\left | \frac{3}{2}^{-},j=2\right \rangle $.

According to the calculation of $\Sigma_c$ in Table~\ref{3-11}, the mass difference of the five $\lambda$-mode states is very small, so it is difficult to distinguish these states well only from the mass spectrum. Combined with a decay width perspective,  The $\Sigma_{c2}(\frac{5}{2}^-)$ state has a mass of 2809 MeV closest to the experimental mass of 2800 MeV, and its decay width is 48.13 MeV, which is very close to the experimental value, so we assume that $\Sigma_c(2800)$ is a good candidates of $\lambda$-mode $J^P=\frac{5}{2}^-$ 1P-wave state, which is consistent with the results of Ref.~\cite{WangWangKaiLei:2021kdd}.

%the experimental decay width of $\Sigma_c(2800)$ is between 62 MeV and 75 MeV. In our calculation, the width of the $\Sigma_{c0}(\frac{1}{2}^-)$ state is 52.47 MeV, which is close to the experimental value, but the mass of the $\Sigma_{c0}(\frac{1}{2}^-)$ state is much smaller than that of the other four states, so we consider that it is unreasonable to regard it as the $\Sigma_c(2800)$ state.

%Sigma_c(2765)
$\mathit{\Sigma_c(2765)}$: As mentioned in section A, based on the provided search results, there is no direct evidence or confirmation that $\Lambda_c(2765)$ is a $\Sigma_c$ state or $\Lambda_c$ state. In this section we discuss $\Sigma_c(2765)$ as a $\Sigma_c$ state in Table\ref{3-11}. For 1P-wave state, the masses of two $\rho$-modes $\Sigma_c$ state is high, 2839 MeV and 2849 MeV, respectively. The mass of the five $\lambda$-modes is very close to the experimental value, and the difference is not more than 30 MeV. From the point of view of decay width, the width value of $\Sigma_{c0}(\frac{1}{2}^-)$ state is closest to the experimental value 52.47 MeV. Therefore, we identify the $\Sigma_c(2765)^+$ as the $\Sigma_{c0}(\frac{1}{2}^-)$ state , which is consistent with the prediction of Ref.~\cite{Chen:2016iyi}.

\begin{table}[h]
\caption{The quantum number of 1S, 2S and 1P-wave $\Sigma_b$ baryons}
\label{3-12}
\begin{tabular}{cccccccccc}
\hline \hline
&Assignments~ &$J$~ &$J_l$~ &$n_\lambda$~ &$L_\rho$~ &$L_\lambda$~ &$L$~ &$S_\rho$~\\ \hline
&$\Sigma_b(\frac{1}{2}^+)$~ &$\frac{1}{2}$~ &0~ &0~ &0~ &0~ &0~ &0~ \\
&$\Sigma_b(\frac{3}{2}^+)$~ &$\frac{1}{2}$~ &1~ &0~ &0~ &0~ &0~ &1~ \\
&$\Sigma_b(2S)$~ &$\frac{1}{2}$~ &0~ &1~ &0~ &0~ &0~ &0~ \\
&$\Sigma_b^*(2S)$~ &$\frac{3}{2}$~ &1~ &1~ &0~ &0~ &0~ &1~ \\
\hline
&$\Sigma_{b0}(\frac{1}{2}^-)$~ &$\frac{1}{2}$~ &0~ &0~ &0~ &1~ &1~ &1~ \\
&$\Sigma_{b1}(\frac{1}{2}^-)$~ &$\frac{1}{2}$~ &1~ &0~ &0~ &1~ &1~ &1~ \\
&$\Sigma_{b1}(\frac{3}{2}^-)$~ &$\frac{3}{2}$~ &1~ &0~ &0~ &1~ &1~ &1~ \\
&$\Sigma_{b2}(\frac{3}{2}^-)$~ &$\frac{3}{2}$~ &2~ &0~ &0~ &1~ &1~ &1~ \\
&$\Sigma_{b2}(\frac{5}{2}^-)$~ &$\frac{5}{2}$~ &2~ &0~ &0~ &1~ &1~ &1~ \\
&$\tilde{\Sigma}_{b1}(\frac{1}{2}^-)$~ &$\frac{1}{2}$~ &1~ &0~ &1~ &0~ &1~ &0~ \\
&$\tilde{\Sigma}_{b1}(\frac{3}{2}^-)$~ &$\frac{3}{2}$~ &1~ &0~ &1~ &0~ &1~ &0~ \\
\hline \hline
\end{tabular}
\end{table}

\begin{table}[h]
\caption{Masses and decay widths of 2S and 1P-wave $\Sigma_b$ states. unit: MeV}
\label{3-13}
\begin{tabular}{ccccccccc}
\hline \hline
&$\Sigma_{bJ_l}(J^P)$~ &Mass~ &$\Lambda_b\pi$~ &$\Sigma_b\pi$~ &$\Sigma_b^*\pi$~ &$\Gamma_{total}$~ \\ \hline
&$\Sigma_{b}(\frac{1}{2}^+,2S)$~ &6167~ &0.62~ &14.79~ &7.84~ &23.25~ \\
&$\Sigma_{b}^*(\frac{3}{2}^+,2S)$~ &6178~ &2.86~ &2.40~ &14.03~ &19.29~ \\ \hline
&$\Sigma_{b0}(\frac{1}{2}^-,1P)$~ &6052~ &50.36~ &$\setminus$~ &$\setminus$~ &50.36~ \\
&$\Sigma_{b1}(\frac{1}{2}^-,1P)$~ &6058~ &$\setminus$~ &150.68~ &0.37~ &151.05~ \\
&$\Sigma_{b1}(\frac{3}{2}^-,1P)$~ &6072~ &$\approx$ 0.0~ &0.68~ &138.72~ &139.40~ \\
&$\Sigma_{b2}(\frac{3}{2}^-,1P)$~ &6090~ &30.78~ &2.06~ &1.12~ &33.96~ \\
&$\Sigma_{b2}(\frac{5}{2}^-,1P)$~ &6103~ &33.75~ &1.28~ &2.60~ &37.63~ \\
&$\tilde{\Sigma}_{b1}(\frac{1}{2}^-,1P)$~ &6168~ &$\setminus$~ &157.09~ &16.23~ &173.32~ \\
&$\tilde{\Sigma}_{b1}(\frac{3}{2}^-,1P)$~ &6171~ &$\approx$ 0.0~ &12.47~ &166.43~ &178.90~ \\
\hline \hline
\end{tabular}
\end{table}

\begin{table*}[ht]
\centering
\caption{Summary of calculations}
\label{4-1}
\setlength{\tabcolsep}{14pt}
\resizebox{\textwidth}{!}{
\begin{tabular}{ccc|cc|cc|ccccc}
\hline \hline
\cline{1-8}&\multicolumn{2}{c|}{$\Lambda_c$} & \multicolumn{2}{c|}{$\Lambda_b$} & \multicolumn{2}{c|}{$\Sigma_c$} & \multicolumn{2}{c}{$\Sigma_b$} \\
&State~  &Candidate~  &State~  &Candidate~  &State~  &Candidate~ &State~  &Candidate~ \\ \hline
&$\Lambda_c(2595)$~ &${\frac{1}{2}}^-(1P)_{\lambda}$~ &$\Lambda_b(5912)$~ &${\frac{1}{2}}^-(1P)_{\lambda}$~ &$\Sigma_c(2800)$~ &${\frac{5}{2}}^-(1P)_{\lambda}$~ &$\Sigma_b(6097)$~ &${\frac{3}{2}}^-/{\frac{5}{2}}^-(1P)_{\lambda}$~ \\
&$\Lambda_c(2625)$~ &${\frac{3}{2}}^-(1P)_{\lambda}$~ &$\Lambda_b(5920)$~ &${\frac{3}{2}}^-(1P)_{\lambda}$~ &$\Sigma_c(2765)$~ &${\frac{1}{2}}^-(1P)_{\lambda}$~ &~ &~ \\
&$\Lambda_c(2765)$~ &${\frac{1}{2}}^+(2S)$~ &$\Lambda_b(6072)$~ &${\frac{1}{2}}^+(2S)$/ ${\frac{1}{2}}^-(1P)_{\lambda}$~ &~ &~ &~ &~\\
&$\Lambda_c(2860)$~ &${\frac{1}{2}}^-(2P)_{\lambda}$~ &$\Lambda_b(6146)$~ &?~ &~ &~ &~ &~\\
&$\Lambda_c(2880)$~ &?~ &$\Lambda_b(6152)$~ &?~ &~ &~ &~ &~ \\
&$\Lambda_c(2910)$~ &${\frac{1}{2}}^-(2P)_{\lambda}$/${\frac{3}{2}}^-(1P)_{\rho}$~ &~ &~ &~ &~ &~ &~\\
&$\Lambda_c(2940)$~ &${\frac{3}{2}}^-(2P)_{\lambda}$~ &~ &~ &~ &~ &~ &~\\
\hline \hline
\end{tabular}}
\end{table*}

\subsection{$\Sigma_b$}

For $\Sigma_b$ baryons, the lowest S-wave states $\frac{1}{2}^+$ and $\frac{3}{2}^+$ haven been observed and confirmed. For the excited state, $\Sigma_b$ experiments only observed $\Sigma_b(6097)$, which is similar to $\Sigma_c(2800)$, and most theorists identify it as a 1P-wave $\lambda$-mode excited state. In this work, we also discuss the $\Sigma_b$ as the 2S and 1P states, the quantum numbers and masses of the P-wave and 2S-wave states are listed in Table \ref{3-12}and the theoretical masses and decay widths are shown in Table \ref{3-13}

$\mathit{\Sigma_b(6097)}$: The mass and width of $\Sigma_b(6097)$ measured experimentally are $6095.8\pm1.7\pm0.4$ MeV and $31.0\pm5.5\pm0.7$ MeV~\cite{LHCb:2018haf}. In Ref.~\cite{Kakadiya:2021jtv}, hypercentral Constituent Quark Model(hCQM) method is used to calculate the energy spectrum, strong decay and electromagnetic decay of $\Sigma_b$. They believes that $\Sigma_b(6097)$ can be interpreted as the $\frac{3}{2}^-$ or $\frac{5}{2}^-$ state of 1P-wave state. Similarly, according to the $^3P_0$ model calculated in Ref.~\cite{Yang:2018lzg}, it is most reasonable to interpret $\Sigma_b(6097)$ as a 1P-wave $\frac{3}{2}^-$ or $\frac{5}{2}^-$ state. Similarly, $\Sigma_b$ also has the problem of mixing, Ref.\cite{Liang:2019aag} studies the variation of its decay width with the mixing angle, and they results show that there is a mixing of $\Sigma_{b1}(\frac{3}{2}^-)$ and $\Sigma_{b2}(\frac{3}{2}^-)$ states in $\Sigma_b(6097)$ at about $20^{\circ}$.

%In the work of Ref~\cite{Chen:2018vuc}, not only the ground state $\Lambda_b$ and $\Sigma_b$, but also the excited state of $\Lambda_b(5912)$ and $\Lambda_b(5920)$ were considered in final state of decay, and the conclusions they obtained were consistent with Ref.~\cite{Yang:2018lzg,Chen:2018vuc}.

%Sigma_b(6097)
In our work, from the perspective of energy spectrum, the theoretical values of 2S-wave and 1P-wave $\rho$-mode are about 100 MeV larger, all of which are interpreted as $\Sigma_b(6097)$ unreasonable, and the masses of 1P-wave $\lambda$-modes $\Sigma_{b2}(\frac{3}{2}^-)$ and $\Sigma_{b2}(\frac{5}{2}^-)$ are much closer, less than 10 MeV difference from the experimental values. From the decay width perspective, the total decay widths of the $\Sigma_{b2}(\frac{3}{2}^-)$ and $\Sigma_{b2}(\frac{5}{2}^-)$ states are 33.96 MeV and 37.63 MeV, which are very close to the experimental values, thus it is reasonable to interpret $\Sigma_{b2}(\frac{3}{2}^-)$, $\Sigma_{b2}(\frac{5}{2}^-)$ as $\Sigma_b(6097)$.\\

%$\Sigma_b(\frac{1}{2}^+)$ and $\Sigma_{b}(\frac{3}{2}^+)$ decay widths of 2S-wave state and $\Sigma_{b0}(\frac{1}{2}^-)$, $\Sigma_{b2}(\frac{3}{2}^-)$ and $\Sigma_{b2}(\frac{5}{2}^-)$ decay widths of 1P-wave are close to the experimental values, and the decay modes of these three P-wave states are dominated by $\Lambda_b \pi$. According to the calculation of the energy spectrum given before,

%Sigma_b(6070)
$\mathit{\Sigma_b(6070)}$: In the previous discussion of $\Lambda_b(6070)$, the decay width of $\Lambda_b(6070)$ as 2S-wave of $\Lambda_b$ state is too narrow compared to experimental values, which is consistent with the conclusion in Ref.~\cite{Liang:2020kvn}. Their work also discuss $\Lambda_b(6070)$ as the $\Sigma_b$ state and consider its mixing angle. In this work, we also consider the possibility of $\Lambda_b(6070)$ as the $\Sigma_b$ state. From Table \ref{3-13}, the mass of $\Sigma_{b0}(\frac{1}{2}^-,1P)$ is 6050 MeV and the decay width is 50.36 MeV, which is very close to the experimental value of $\Lambda_b(6070)$. So we prefer to interpret $\Lambda_b(6070)$ as a $\Sigma_b$ state with a quantum number of $\frac{1}{2}^-$.

\section{Summary}

In this work, we simultaneously investigate the mass spectrum and decay widths of singly heavy baryons in the same theoretical framework, aiming to achieve a more self-consistent study.
We calculate the energy spectrum of the excited state of the singly heavy baryon by applying the chiral quark model under $SU(3)$ symmetry, taking into account the effects of
spin-orbit couplings and the tensor force. Afterwards we bring the theoretical masses and eigenvalues into the $^{3}P_{0}$ model to obtain the width of the strong decay of
the singly heavy baryon, and we take into account not only the two-body strong decay but also the three-body decay. The energy spectra and decay widths of $\Lambda_{c(b)}$
and $\Sigma_{c(b)}$ excited states are obtained relatively self-consistently by applying a complete set of parameters, which are in good agreement with many experimentally
newly discovered heavy baryon states. We have summarised the results in Table \ref{4-1}.

A significant challenge in our study is achieving consistent descriptions of both the mass spectrum and decay widths with one unified set of parameters.
Due to the uncertainty of the model itself and the experimental data, our results cannot be completely consistent with the experimental data.
But most of our results are still in good agreement with the experimental values within a certain margin of error allowed, hoping to provide a reference for the experiment.
In this work, we only study the properties of $2S$-wave, $1P$-wave and $2P$-wave $\Lambda_{c(b)}$ and $\Sigma_{c(b)}$ baryons.
In the future, we will extend the study to other heavy baryons, not only considering P-wave excitations, but also higher wave excitations.
Regarding three-body decays, we have only proposed a simplified approach in this work, and further refinements to the three-body decay process will be pursued in future studies, and we hope to get a systematic work on both the energy spectrum and decay widths of heavy baryons.

\acknowledgments{This work is supported partly by the National Natural Science Foundation of China under Contracts Nos. 11675080, 11775118, 12305087, 11535005 and 11865019.}

\end{document}